\documentclass[lettersize,journal,onecolumn]{IEEEtran}
\usepackage{amsfonts}
\usepackage{amsmath}
\usepackage{amsmath}
\usepackage{amsthm}
\newtheorem{assumption}{Assumption}
\usepackage{algorithmic}
\usepackage{algorithm}
\usepackage{subcaption}
\usepackage{makecell}
\usepackage[normalem]{ulem}
\usepackage{array}
\usepackage{subfig}
\usepackage{textcomp}
\usepackage{amssymb}
\usepackage{caption}
\usepackage{xcolor}
\usepackage{stfloats}
\usepackage{url}
\usepackage{verbatim}
\usepackage{graphicx}
\usepackage{cite}

\begin{document}

% \title{Cell-Free~Integrated~HAPS-Terrestrial~Networks: A~Low~Complexity~Large~Scale Distributed Interference Management}

\title{Two-Level Distributed Interference Management for Large-Scale HAPS-Empowered~vHetNets}
\author{Afsoon Alidadi Shamsabadi,~\IEEEmembership{Senior Member,~IEEE,} Animesh Yadav,~\IEEEmembership{Senior Member,~IEEE,} and~Halim~Yanikomeroglu,~\IEEEmembership{Fellow,~IEEE}
\thanks{Afsoon Alidadi Shamsabadi and Halim Yanikomeroglu are with the Department of Systems and Computer Engineering, Carleton University, Ottawa, ON K1S 5B6, Canada (e-mail: afsoonalidadishamsa@sce.carleton.ca; halim@sce.carleton.ca). Animesh Yadav is with the School of Electrical Engineering and Computer Science, Ohio University, Athens, OH 45701 USA (e-mail: yadava@ohio.edu).}
}
\maketitle
\begin{abstract}
High altitude platform stations (HAPS) offer a promising solution for achieving ubiquitous connectivity in next-generation wireless networks (xG). Integrating HAPS with terrestrial networks, creating HAPS-empowered vertical heterogeneous networks (vHetNets), significantly improves coverage and capacity and supports emerging novel use cases. In HAPS-empowered vHetNets, HAPS and terrestrial network tiers can share the same spectrum, forming harmonized spectrum vHetNets that enhance spectral efficiency (SE). However, harmonized spectrum vHetNets face major challenges, including severe co-channel interference and scalability in large-scale deployments. To address the first challenge, we adopt a cell-free multiple-input multiple-output (MIMO) network architecture in which users are simultaneously served by multiple base stations using beamforming. However, beamforming weight design leads to a nonconvex, high-dimensional optimization problem, highlighting the scalability challenge. To address this second challenge, we develop a two-level distributed proportional fairness beamforming weight design (PFBWD) algorithm. This algorithm combines the augmented Lagrangian method (ALM) with a three-block ADMM framework. Simulation results demonstrate the performance improvements achieved by integrating HAPS with standalone terrestrial networks, as well as the reduced complexity and signaling overhead of the distributed algorithm compared to centralized algorithms.
\end{abstract}
\section{Introduction}
The next-generation (xG) wireless networks are evolving toward artificial intelligence (AI)-driven, sensing-based systems that must deliver immersive throughput, hyper reliability, and low latency, while ensuring connectivity for everyone worldwide~\cite{IMT-2030}. A wide range of innovative technologies is under consideration to meet these goals, spanning multiple layers of the telecommunication network, including terahertz frequency bands, movable antenna architectures, holographic beamforming, semantic communication, and advanced AI-driven resource management strategies. Meeting the ambitious performance and coverage demands requires more than just novel technologies. This also demands innovative network architectures, coupled with efficient performance optimization algorithms that can fully unlock their potential~\cite{6GVTM}. Among these architectural innovations, non-terrestrial networks (NTN), including space satellites, high altitude platform stations (HAPS), and uncrewed aerial vehicles, represent a particularly promising direction~\cite{NTNApplications}. Unlike earlier deployments that primarily targeted rural and remote areas, future NTN will be tightly integrated with urban terrestrial networks to provide global coverage with seamless quality of experience.

\textcolor{black}{In future xG wireless networks, HAPS is expected to play a critical role, particularly in urban environments with high user density and traffic demand. These roles span a wide range of use cases, from enabling aerial mobile edge computing to providing connectivity for ground users and Internet of Things (IoT) devices~\cite{ZiyeMEC,ZiyeJia}.} Positioned quasi-stationary in the stratosphere layer of the Earth at an altitude of approximately 20 km above the ground, HAPS offers several advantages over both space-based satellites and terrestrial networks. For instance, HAPS provides lower latency compared to space satellites, greater flexibility to support advanced technologies (e.g., ultra massive multiple-input multiple-output (umMIMO), reconfigurable intelligent surfaces), and a wider coverage area than terrestrial macro base stations (MBSs). As such, HAPS can serve as a complementary platform with the potential to help meet the demanding requirements of future networks, in conjunction with legacy terrestrial and satellite systems~\cite{Our-WCL}. Moreover, the large surface area of HAPS enables the integration of ambient energy harvesting resources, contributing to the development of a more sustainable network architecture~\cite{HAPSSurvey,HAPS-Alouini}. Referred to by the International Telecommunication Union (ITU) as HAPS as International Mobile Telecommunications (IMT) base stations (HIBS), HAPS can be integrated with existing terrestrial networks in urban environments to enhance coverage and capacity, thereby supporting the fulfillment of future network KPIs such as ubiquitous connectivity, immersive data rate, and sustainability. Such integrated HAPS-terrestrial networks are referred to as HAPS-empowered vertical heterogeneous networks (vHetNets)~\cite{IM_Magazine}.

Various tiers in vHetNets can operate over the same or different frequency bands~\cite{SpecSharing}. Notably, the World Radiocommunication Conference 2023 (WRC-23) allocated additional IMT-identified spectrum to HAPS~\cite{itu_wrc}. This development is pivotal for enabling the integration of HAPS with terrestrial wireless networks under a harmonized spectrum framework, where both HAPS and terrestrial tiers share the same frequency band. While such spectrum sharing enhances spectral efficiency~(SE), it also leads to performance degradation due to inter- and intra-tier interference propagation~\cite{IM_Magazine}. This challenge is further amplified by the wide coverage area of HAPS. The high altitude of HAPS enables vHetNets to provide wide-area coverage that includes multiple terrestrial MBSs. Consequently, effective coordination between HAPS and these MBSs is essential for delivering optimal service to user equipment (UEs). Unlike terrestrial MBSs, which primarily coordinate with their immediate neighbors, HAPS must interact with all MBSs within its coverage area. This requirement results in large-scale optimization problems with greater complexity, often making rapid convergence difficult to achieve. Moreover, efficient decision-making depends on exchanging network status information between MBS and HAPS, which introduces significant signaling overhead on communication links.
\vspace{-2mm}
\subsection{Related works}
Since the wireless channel is interference-limited, designing efficient interference management strategies has a long research history~\cite{InterferenceManagement}. Recent studies have explored spectrum sharing and interference suppression techniques in integrated HAPS-terrestrial networks. To this end, the authors in \cite{SpectrumSharing} proposed an interference canceler and coordination mechanism to facilitate spectrum sharing between HAPS and terrestrial systems. In \cite{HAPSIM-Cell}, the authors introduced a cell design method for HAPS aimed at extending coverage while maintaining coexistence with terrestrial mobile networks. In addition,~\cite{HAPSIM-TVT} presented a codebook-based interference suppression approach for space-air-ground integrated networks, demonstrating improved performance in managing inter-tier interference. In our previous work~\cite{Our-WCL}, we developed an interference management algorithm for an integrated HAPS-terrestrial network, leveraging subcarrier allocation and power control. More recently, the authors in~\cite{HuseyinArslan} proposed an interference mitigation framework to enable coexistence between NTN and terrestrial networks.

Among the existing approaches for managing interference, beamforming weight design has emerged as an efficient strategy~\cite{BFforIM}. To this end, in~\cite{our-CL} and \cite{our-ICC}, we developed three centralized interference management algorithms aimed at optimizing user association and beamforming weights under different objective functions: weighted sum rate (WSR), max-min fairness (MMF), and proportional fairness (PF). The performance of the proposed algorithms was subsequently evaluated in both vHetNets and standalone terrestrial network scenarios. Furthermore, the authors in \cite{HetNetBF} formulated a max-min sum rate beamforming design problem in a HetNet with hardware impairments. In~\cite{REBF}, a centralized beamforming weight design algorithm was proposed for cell-free terrestrial networks to optimize the tradeoff between energy efficiency and SE. However, due to the high computational and signaling overheads associated with centralized algorithms, significant efforts have also been devoted toward developing distributed algorithms.

Accordingly, in~\cite{multi-groupBF}, the authors introduced an alternating direction method of multipliers (ADMM)-based beamforming approach for multi-group multicast systems under two objective functions of minimizing the total transmit power and MMF. In~\cite{StochasticADMM}, the authors developed a stochastic ADMM framework for coordinated multi-cell beamforming in a smartgrid powered coordinated multicell downlink system. In addition, the authors in~\cite{ADMMforDist} proposed an ADMM-based distributed beamforming algorithm to minimize total transmit power under signal-to-interference-and-noise ratio (SINR) constraints in multi-cell networks. Moreover, \cite{ADMMCellFree} applied ADMM to power-efficient downlink beamforming in cell-free mMIMO, also targeting SINR-constrained optimization across distributed access points. 
In~\cite{APS_BF2026}, the authors proposed a deep reinforcement learning (DRL)-based beamforming approach for mMIMO aerial communications, aiming at maximizing the sum rate.

Although the two-block ADMM approach has been widely adopted in distributed algorithm design~\cite{BoydADMM}, its limitations hinder its applicability to general nonconvex problems. To address this, more sophisticated methods have been explored. In~\cite{FullDimension}, the authors formulated a joint user association and beamforming weight design problem aiming at maximizing the ergodic sum rate. Accordingly, the SINR term has been approximated with the average signal-to-leakage-and-noise ratio, and a three-step Gaussian belief propagation-based distributed solver is proposed to solve the problem. In~\cite{DistributedBFBP}, the authors formulated a sum rate maximization problem and introduced a belief propagation algorithm for cooperative MIMO systems that enables efficient downlink beamforming through decentralized message passing. In recent years, deep learning-based approaches have also gained traction. The authors in \cite{DRLCellFreeDist} employ DRL for distributed uplink beamforming in cell-free networks, considering a sum rate objective function. Similarly,~\cite{GraphUABF} uses distributed graph-based learning to jointly optimize user association and beamforming in multi-cell networks to maximize the sum rate. Collectively, these contributions underscore the growing emphasis on distributed beamforming frameworks for xG wireless networks. Existing distributed studies are primarily developed for terrestrial networks operating at relatively smaller scales compared to HAPS-empowered vHetNets. These works typically focus on sum rate maximization, which often prioritizes overall throughput at the expense of the minimum achievable SE. Moreover, to address the inherent nonconvexity of the formulated optimization problems, existing approaches rely on simplifying approximations or relaxations, which can limit the achievable performance. Consequently, distributed solutions that jointly address scalability, fairness guarantees, and harmonized-spectrum operation in cell-free HAPS-empowered vHetNets remain largely unexplored.
\vspace{-3mm}
\subsection{Motivations and Contributions}
It is evident that centralized algorithms favor small-scale systems; hence, they are practically infeasible in harmonized spectrum HAPS-empowered vHetNets, which are large-scale systems for several reasons. First, the large coverage area of HAPS encompasses a large number of MBSs, serving a large number of UEs. Since HAPS and MBSs are sharing the same spectrum, they need to coordinate with each other in order to serve UEs. Considering the large coverage area of HAPS, this creates a large-scale network. Second, the use of ultra mMIMO involves a large number of antenna elements, particularly at HAPS with a large surface area, resulting in a high-dimensional optimization problem. Therefore, the development of a distributed algorithm for beamforming design becomes necessary for handling large-scale networks such as HAPS-empowered vHetNets. Furthermore, centralized algorithms must collect all network information, including channel coefficients, solve the optimization problem, and distribute the results to the base stations (BSs\footnote{The term BS in this work collectively refers to HAPS and MBSs.}). However, exchanging data between HAPS and MBSs, considering the positioning of HAPS, imposes additional latency to the network. Therefore, distributed algorithms to design the important network parameters are a key solution to overcome these challenges.

In addition to the aforementioned challenges, the formulated problems in vHetNets are inherently nonconvex, especially in interference-dominant networks~\cite{IM_Magazine}. To address this, various reformulation linearization techniques (RLT) are often employed to approximate the original problem and solve it iteratively through successive convex approximation (SCA)~\cite{SCA}. However, such iterative methods require solving multiple optimization problems iteratively, making them challenging for real-time or low-latency applications in dynamic networks. This reinforces the necessity of developing distributed algorithms tailored for nonconvex interference management problems in vHetNets. However, directly applying conventional distributed approaches such as two-block ADMM to general nonconvex constrained problems does not guarantee convergence unless relaxation schemes are used \cite{jiang2019structured}. To this end, multiple variants of ADMM approaches have been developed to address the nonconvexity of the optimization problems~\cite{ADMMVariants}. Accordingly, the authors in \cite{TWoStepDistributed} proposed a two-level distributed algorithm that can solve the nonconvex constrained problem in a distributed manner, with guaranteed convergence. Specifically, they introduce a key reformulation that guarantees the conditions for convergence of multi-block ADMM algorithms. The resulting two-level algorithm embeds a specially structured three-block ADMM in the inner loop within an augmented Lagrangian method (ALM) framework~\cite{TWoStepDistributed}. This algorithm is selected for its demonstrated advantages in accelerating convergence and improving feasibility in challenging nonconvex settings. Its ability to handle multi-block constraints robustly makes it well-suited for the problem structure.

In this work, we consider a harmonized spectrum HAPS-empowered vHetNet, where HAPS and MBSs are serving UEs in an overlapped coverage area. This novel network architecture is a potential solution to improve coverage and capacity in urban areas. In the system model, the BSs are equipped with MIMO antenna arrays, empowering the BSs to serve UEs using beamforming. Our earlier works investigated related but fundamentally different system models. In \cite{our-CL}, a centralized algorithm was developed for joint user association and beamforming, where each UE was served by only one BS (either the HAPS or an MBS). The follow-up study in \cite{our-ICC} extended the analysis to multiple objective functions (MMF, PF, and WSR) but still relied on a centralized optimization framework and a single-BS association model. These studies revealed two important insights: (i) allowing UEs to be served by multiple BSs can significantly improve spectral efficiency~\cite{our-CL}, and (ii) PF provides a desirable balance between maximizing the sum data rate and the minimum SE of the network. Motivated by these observations, in this paper, we adopt a cell-free architecture in which UEs are jointly served by MBSs and HAPS, simultaneously. Despite recent advances in cell-free architectures and distributed optimization, how to enforce proportional fairness in harmonized-spectrum vHetNets under severe inter-tier interference, without centralized channel state information (CSI) aggregation, remains an open problem. 

We formulate a PF beamforming weight design (PFBWD) problem tailored to this architecture. Unlike \cite{our-CL,our-ICC}, which rely on centralized optimization and single-BS association, this work addresses a cell-free HAPS-empowered vHetNet architecture under a harmonized spectrum, where centralized CSI collection is infeasible. Moreover, while \cite{TWoStepDistributed} proposes a generic two-level distributed framework, this paper makes a nontrivial contribution by reformulating a highly coupled PFBWD problem in HAPS-empowered vHetNets into a structure that satisfies the convergence conditions of \cite{TWoStepDistributed}, which is not straightforward due to the interference coupling and network-side fairness objective function. To this end, we reformulate the original problem into a distributed structure and develop a two-level distributed algorithm in which each BS optimizes its local variables using only local channel coefficients, without exchanging channel coefficients.

Accordingly, the main contributions of this paper are summarized as follows:
\begin{itemize}
\item We study a harmonized spectrum HAPS-empowered vHetNet with overlapping HAPS–terrestrial coverage under a cell-free architecture, where UEs are jointly served by multiple MBSs and the HAPS. In this setting, we formulate a PFBWD optimization problem that explicitly captures the strong inter-tier interference coupling inherent to such networks.

\item We show that the resulting PFBWD problem cannot be efficiently solved using centralized optimization due to its large scale, dense interference structure, and the impracticality of global channel coefficient collection in HAPS-empowered vHetNets. To address this, we reformulate the problem into a distributed structure that enables BS-level optimization using only local CSI, without inter-BS channel exchange.

\item Based on the proposed reformulation, we develop a two-level distributed optimization framework in which a structured three-block ADMM is embedded within an outer ALM. The proposed framework is tailored to the considered cell-free HAPS-empowered vHetNet and is capable of handling the nonconvex formulated PFBWD problem.

\item We establish the convergence of the proposed two-level distributed algorithm and validate its performance for large-scale nonconvex interference management through simulations.
\end{itemize}

\color{black}
\textit{Nomenclature:} We use italic lowercase letter $x$ for scalars, bold lowercase letter $\mathbf{x}$ for vectors, bold uppercase letter $\mathbf{X}$ for matrices, and calligraphic uppercase letter $\mathcal{X}$ for sets.
The $i$th entry of vector $\mathbf{x}$ is denoted by $\mathbf{x}_i$, and the $(i,j)$ entry of matrix $\mathbf{X}$ by $\mathbf{X}_{i,j}$. The notation $\|\cdot\|$ represents the Euclidean norm, while $\|\cdot\|_F$ denotes the Frobenius norm. The Kronecker product is represented by $\otimes$. The superscript $(\cdot)^\top$ and $(\cdot)^H$ denote transpose and Hermitian transpose, respectively.

The remainder of the paper is organized as follows. Section~\ref{Sec:model} introduces the system model and formulates the \mbox{PFBWD} optimization problem. Section~\ref{Sec:Reformulation} provides a distributed reformulation of the original nonconvex problem. Section~\ref{Sec:DistributedAlgorithm} describes the developed approach and algorithm. Section~\ref{Sec:Complexity} discusses the convergence properties and computational complexity of the algorithm. Section~\ref{Sec:Results} presents and analyzes the numerical results, and finally, Section~\ref{Sec:Conclusion} concludes the paper.
\begin{figure}[t]
    \centering
    \captionsetup{justification=centering}
    \includegraphics[width=0.75\linewidth]{\detokenize{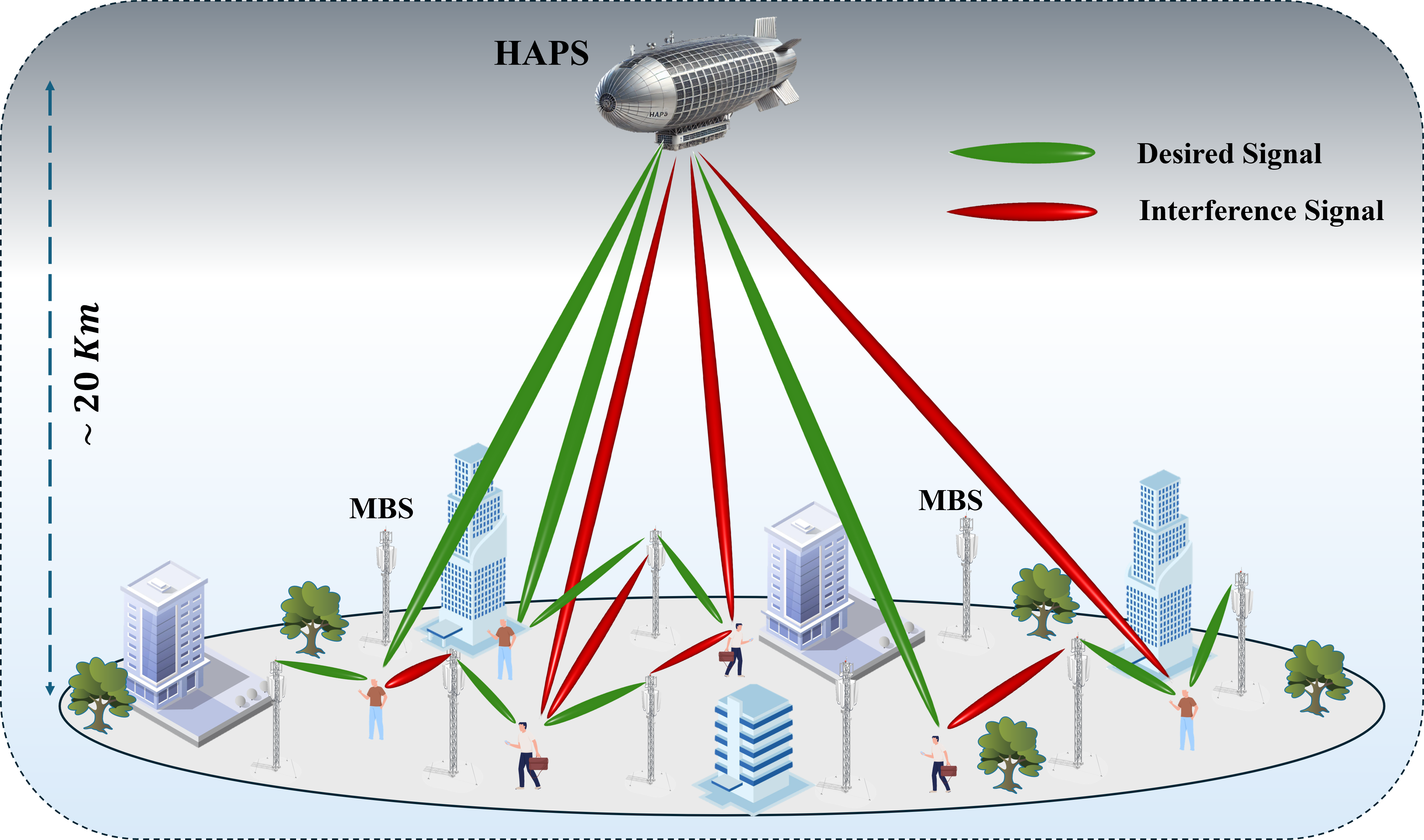}}
    \caption{\small HAPS-empowered vHetNet system model.}
    \label{SystemModel}
\end{figure}
\section{System Model and Problem Formulation}\label{Sec:model}
We consider a cell-free HAPS-empowered vHetNet consisting of one HAPS and $B$ MBSs, simultaneously serving $U$ single antenna UEs as illustrated in Fig.~\ref{SystemModel}. The vHetNet operates in the downlink channel within the sub-6 GHz frequency band. UEs and BSs are indexed by $u \in \mathcal{U} \triangleq \{1,\ldots,U\}$, and $b \in \mathcal{B} \triangleq \{1,\ldots,B+1\}$, respectively, where index $b = B+1$ is reserved for the HAPS. The MBSs and HAPS are equipped with MIMO antennas with a total number of $N_b=N^H_b \times N^V_b$ antenna elements in a uniform planar array configuration, where $N^H_b$ and $N^V_b$ are the number of antenna elements in the horizontal and vertical axes, respectively. To efficiently manage interference, BSs employ beamforming to communicate with UEs using user-specific beams. Therefore, the proper design of beamforming weights becomes essential and is the primary focus of this work. We denote the beamforming weight matrix at BS $b$ by $\mathbf{W}^b = [\mathbf{w}^b_1, \dots, \mathbf{w}^b_U] \in \mathbb{C}^{N_b \times U}$, where $\mathbf{w}^{b}_{u} \in \mathbb{C}^{N_b}$ refers to the beamforming vector for UE $u$ at BS $b$, and each element $w^b_{r,u}$ represents the complex weight applied to antenna element $r \in \{1,\dots,N_b\}$ when transmitting to UE $u$. Accordingly, we define $\mathbf{H}^b=[\mathbf{h}^b_1,\dots,\mathbf{h}^b_U] \in \mathbb{C}^{N_b \times U}$, as the channel coefficients matrix at BS b, where $\mathbf{h}^b_u=[h^b_{1,u},\dots,h^b_{N_b,u}]^\top \in \mathbb{C}^{N_b}$ denotes the channel vector between BS $b$ and UE $u$. We assume that all UEs share the same time-frequency resources with equal bandwidth allocation. Consequently, interference is propagated from all serving BSs to all UEs. \textcolor{black}{We assume that other terrestrial tiers, such as small cells, operate over an orthogonal spectrum to the considered vHetNet. Therefore, no interference from these tiers is experienced by the UEs.}

Due to the strategic positioning of the HAPS at an altitude of $20$~km above the ground, its role in the network and its interaction with MBSs and UEs are fundamentally different from those of terrestrial BSs, including both MBSs and small-cell BSs (SBSs). In particular, the large elevation angle and strong line-of-sight (LoS) connection between the HAPS and a large number of serving UEs result in channel characteristics that differ significantly from terrestrial links, which are typically dominated by non-LoS (NLoS) propagation and localized interference. As a consequence, HAPS transmissions generate a broad interference footprint that simultaneously impacts a large number of UEs and MBSs over an extended area, necessitating large-scale coordination under harmonized spectrum operation. This behavior is fundamentally different from conventional MBS–SBS deployments, where interference is spatially confined and coordination is typically limited to a small cluster of neighboring cells. Accordingly, we adopt two distinct channel models for the HAPS and terrestrial tiers, which are described in detail in the following subsection.

\subsection{MBS-UE Channel} 
We consider small-scale fading, free-space path loss (FSPL), and shadowing for the channel between each UE and MBS. Particularly, the channel coefficient between the antenna element $r \in \{1,\dots,N_b\} $ of MBS $b$ and UE $u$, denoted as $h^{b}_{r,u}$, can be formulated as follows\cite{MBSChannel}:
\begin{equation}\label{MBSchannelgain}
    h^b_{r,u}=\frac{\overline{h}^b_{r,u}\xi^b_u} {\sqrt{{PL}_{b,u}}},~\forall r,~\forall u,~\forall b,
\end{equation}
where $\overline{h}_{r,u}$ represents the small-scale fading channel coefficient and follows a Rayleigh distribution. $\xi^b_u=10^{{{\xi_u^{'b}}}/10}$ denotes the log-normal shadowing gain, where ${\xi_u^{'b}}$ is the Gaussian random variable with $\mathcal{N}(0, \sigma_\xi\, \text{dB})$. ${PL}_{b,u}$ refers to the FSPL between MBS $b$ and the UE $u$, and can me computed as
\begin{equation}\label{FSPL}
   {{PL}_{b,u}}=\Big(\frac{4\pi f_c d_{b,u}}{c}\Big)^2,~\forall b,~\forall u,
\end{equation}
where $f_c$ is the carrier frequency (in Hz), $d_{b,u}$ (in m) is the distance between the BS $b$ and the UE $u$, and $c$ represents the speed of light in free space.
\subsection{HAPS-UE Channel} 
\begin{figure}[t]
    \centering
    \captionsetup{justification=centering}
    \includegraphics[width=0.5\linewidth]{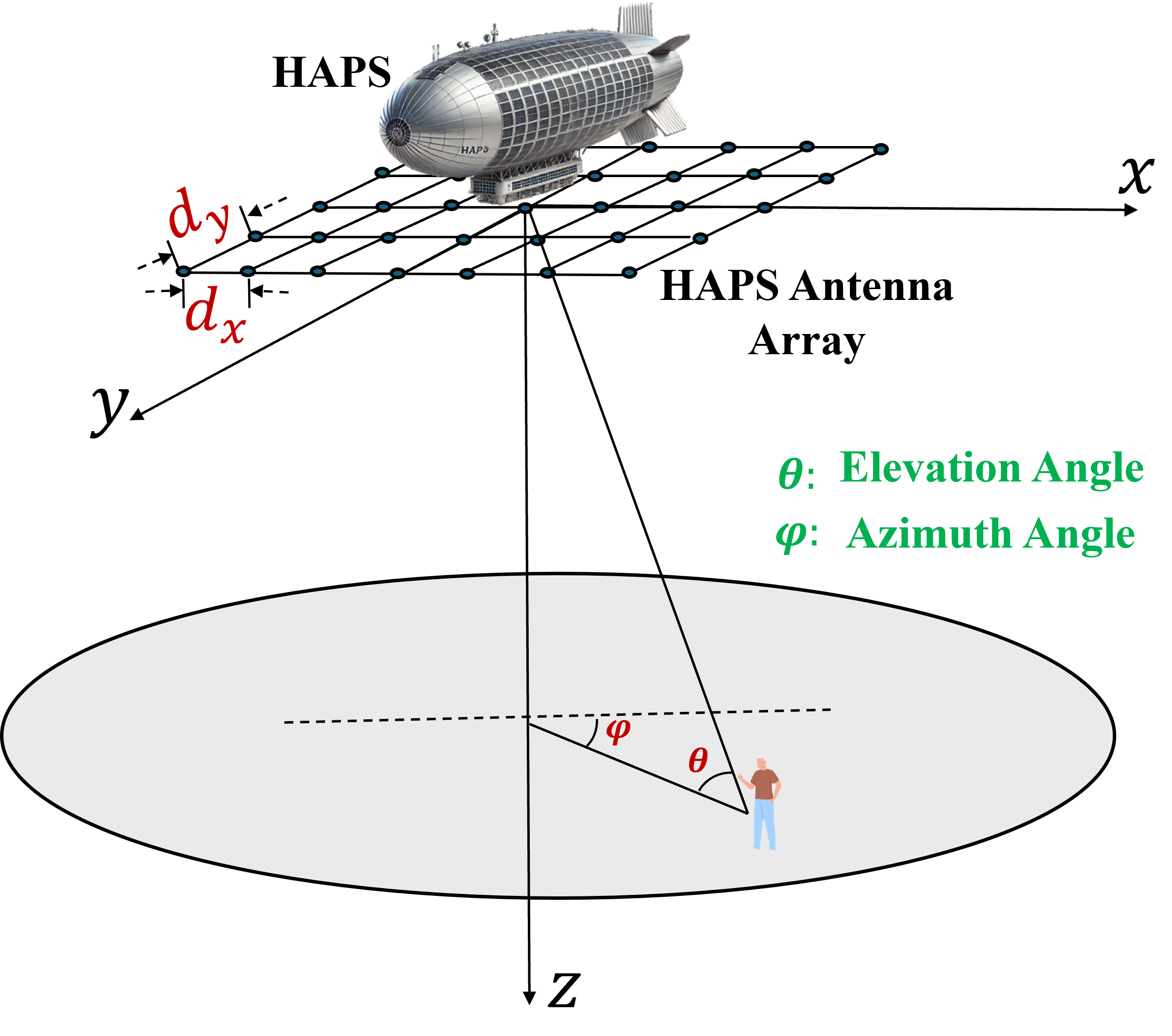}
    \caption{\small HAPS antenna architecture.}
    \label{fig_2}
\end{figure}
Due to the high altitude of HAPS, there exist strong LoS links with a high probability to UEs.  However, small-scale fading may still occur due to the multipath characteristics of urban environments. Particularly, $\mathbf{h}^{B+1}_u$, representing the channel coefficient vector between HAPS and UE $u$, can be modeled as a three-dimensional (3D) Rician fading channel with a dominant LoS and NLoS components as \cite{HAPS-MIMO}
\begin{equation}
    \mathbf{h}_u^{B+1}=\frac{1}{\sqrt{{PL}_{(B+1),u}}}\Big(\sqrt{\frac{1}{1+K_u}}\overline{\mathbf{h}}_u+\sqrt{\frac{K_u}{1+K_u}}\hat{\mathbf{h}}_u\Big),~\forall u,\qquad \qquad \qquad
\end{equation}
where ${PL}_{(B+1),u}$ represents the FSPL between HAPS and UE $u$, calculated according to \eqref{FSPL}. $K_u$ is the Rician factor for UE $u$, and $\overline{\mathbf{h}}_u \in \mathbb{C}^{N_{B+1}}$ represents the NLoS component of the channel vector with its elements from a normal random distribution with zero mean and unit variance, $\mathcal{NC}(0,1)$. Accordingly, $\hat{\mathbf{h}}_u$ is the LoS component of the channel vector, given by \cite{HAPS-MIMO} as
\begin{equation}\label{LoS-HAPS}
    \hat{\mathbf{h}}_u=\mathbf{a}(\theta_u,\phi_u) \otimes \mathbf{b}(\theta_u,\phi_u),~\forall u,
\end{equation}
where
\begin{IEEEeqnarray}{lcl}\label{ab}
    \mathbf{a}(\theta_u,\phi_u)=[1,e^{j2\pi d_h}, \dots, e^{j2\pi (N^\text{H}_{B+1}-1)d_h}]^\top,~\forall u, \IEEEyesnumber \IEEEyessubnumber* \label{a}\\
    \mathbf{b}(\theta_u,\phi_u)=[1,e^{j2\pi d_v}, \dots, e^{j2\pi (N^\text{v}_{B+1}-1)d_v}]^\top,~\forall u, \label{b}
\end{IEEEeqnarray}
where $d_h=d_\text{x} \cos{\theta_u}\sin{\phi_u}/\lambda$ and $d_v=d_\text{y} \cos{\theta_u}\cos{\phi_u}/\lambda$. As depicted in Fig.~\ref{fig_2}, $\theta_u \in [0,\pi/2]$ and $\phi_u \in [-\pi,\pi)$ are the elevation and azimuth angles of UE $u$. Further, $d_\text{x}$ and $d_\text{y}$ are the antenna elements spacing in $y$ and $x$ directions and $\lambda=c/f_c$ represents the wavelength.

As mentioned earlier, we adopt the cell-free as the underlying network architecture where all UEs are served by all BSs. We assume that each BS has perfect information regarding the channel coefficients between antenna elements and UEs. Before defining the SINR expression, we first describe the received signal model under the considered cell-free transmission architecture. In this setup, each UE is jointly served by multiple BSs. \textcolor{black}{Following the standard coherent joint-transmission model~\cite{Ngo2017CellFree}, all serving BSs transmit the same data symbol $s_u$ intended for UE $u$, each precoded by a BS-specific beamforming vector $\mathbf{w}^{b}_{u}$. As a result, the desired signal components transmitted from different BSs add coherently at the receiver, while interference arises only from symbols $s_k$ intended for other UEs $k \in \mathcal{U} \setminus \{u\}$.} The received signal at UE $u$ is
therefore given by
\begin{equation}
    y_u = 
    \underbrace{\sum_{b\in\mathcal{B}} (\mathbf{h}^{b}_{u})^{H}\mathbf{w}^{b}_{u} s_u}_{\text{combined desired signal}}
    +
    \underbrace{\sum_{k \in \mathcal{U} \setminus \{u\}} \sum_{b\in\mathcal{B}}
    (\mathbf{h}^{b}_{u})^{H}\mathbf{w}^{b}_{k} s_k}_{\text{multi-user interference}}
    + n_u,~\forall u, \qquad
\end{equation}
where $(\mathbf{h}^{b}_{u})^{H}$ denotes the channel from BS $b$ to UE~$u$, $\mathbf{w}^{b}_{k} \in \mathbb{C}^{N_b}$ refers to the beamforming vector for UE $k$ at BS $b$, and $n_u$ is the additive white Gaussian noise (AWGN) with variance $\sigma^2_n$. Since all serving BSs transmit the same symbol $s_u$, the UE
does not need to decompose per-BS contributions; it simply decodes a single effective
combined stream. Based on this received signal model, the SINR of UE $u$ in the cell-free vHetNet can be expressed as~\cite{Marzetta}
\begin{equation}\label{eq:SINR}
    \gamma_u=\cfrac{\left|\sum_{b\in \mathcal{B}}\left(\mathbf{h}^{b}_{u}\right)^H \mathbf{w}^{b}_{u}\right|^2}{\sum_{k \in \mathcal{U} \setminus \{u\}}\left|\sum_{b\in \mathcal{B}}{\left(\mathbf{h}^{b}_{u}\right)^H \mathbf{w}^{b}_{k}}\right|^2+\sigma^2_n},~\forall u.
\end{equation}

Since the PF criterion effectively improves both the minimum SE of the UE and the sum SE~\cite{our-ICC}, we adopt it as the objective function in our study. Accordingly, the PF objective is expressed as the sum of logarithms of SEs over all UEs, where the SE of UE~$u$ is defined as $\log_2(1+\gamma_u)$. This objective function follows the classical PF formulation introduced in~\cite{PFCharging}, in which a logarithmic utility function is employed due to its increasing yet concave nature. In this utility, the logarithm naturally promotes fairness by giving more weight to UEs with low SINR and reducing the influence of UEs with already high SINR. Consequently,  the performance of weaker UEs improves, while still maintaining strong overall system performance.

Accordingly, the preliminary PFBWD optimization problem can be expressed as
\begin{IEEEeqnarray*}{lcl}\label{eq:Original_Problem}
    &\underset{\mathbf{W}^b}{\operatorname{maximize}}\,\, & ~\sum_{u\in\mathcal{U}} {\log(\log_2(1+\gamma_u))} \,  \IEEEyesnumber \IEEEyessubnumber* \label{eq:Original_Problem_Obj}\\
    &\text{s.t.} & \gamma_u \geq \gamma_{\text{min}},~\forall u, \qquad \label{eq:Original_Problem_const0}\\
    && {\|\mathbf{W}^b \|^2_F}\leq P^{\text{max}}_b, \, ~ \forall b,\qquad \label{eq:Original_Problem_const1}
\end{IEEEeqnarray*}
where $P_b^{\text{max}}$ is the maximum allowable transmit power of BS $b$, and $\gamma_{\text{min}}$ is the minimum required SINR for each UE. In problem \eqref{eq:Original_Problem}, constraint \eqref{eq:Original_Problem_const0} ensures that each UE $u$ meets its minimum data rate requirement, while constraint \eqref{eq:Original_Problem_const1} restricts the total transmit power of each BS $b$ to not exceed $P_b^{\text{max}}$. Each BS is subject to its own maximum transmit power constraint, i.e., the HAPS and MBSs are assigned distinct power budgets that reflect their hardware capabilities and link budgets. Problem \eqref{eq:Original_Problem} is nonconvex due to the objective function \eqref{eq:Original_Problem_Obj} and the nonconvex nature of constraint \eqref{eq:Original_Problem_const0}, rendering it intractable for conventional solvers. Typically, problems of this nature can be addressed by developing an iterative algorithm using the SCA framework. In each iteration of the SCA, a convex approximation of the original problem is solved. The solution obtained at convergence is considered suboptimal relative to the original nonconvex problem. Additionally, the approximated problems can be solved using either a centralized or a distributed algorithm. However, for large-scale problems, it is advisable to avoid a centralized algorithm, as it requires collecting channel coefficients from all BSs to solve the optimization problem, and thereafter distributing the decisions back to the BSs. This process imposes significant signaling overhead on the network. Additionally, the centralized approach must solve a large-scale optimization problem to determine the decision variables for the entire network, resulting in high computational complexity.

The problem at hand is large in scale and complexity, thereby necessitating the development of a distributed algorithm to solve it efficiently. To enable distributed computation, the original centralized formulation presented in \eqref{eq:Original_Problem} must first be transformed into a distributed formulation to which a distributed optimization method can be applied. The key objective of this reformulation is to enable parallel decomposition, whereby the problem is structured in a way that enforces per-BS constraints. This structure allows the overall problem to be divided and solved across multiple BSs simultaneously. In the subsequent section, we focus on this transformation process, aiming to recast the original problem \eqref{eq:Original_Problem} into a distributed format suitable for distributed implementation. Although the proposed distributed algorithm is general and applicable to any multi-BS deployment, its advantages become particularly pronounced in cell-free architectures, where each BS serves all UEs simultaneously, resulting in a larger optimization dimension and requiring tight coordination across BSs.
\vspace{-2mm}
\section{Distributed Reformulation of the Original Problem \eqref{eq:Original_Problem}} \label{Sec:Reformulation}
In this section, we detail the series of reformulations required to transform the problem \eqref{eq:Original_Problem} to a distributed format that can be solved by the two-level distributed algorithm. To this end, first, we deal with the objective function \eqref{eq:Original_Problem_Obj}. We 
introduce $U$ slack variables $t_u,~\forall u,$ as the lower bound for \mbox{$\log_2(1+\gamma_u),~\forall u$}. Accordingly, the original objective function \eqref{eq:Original_Problem_Obj} can be equivalently replaced by \mbox{$f(\mathbf{t})=\sum_{u=1}^{U} \log_2(t_u)$}, where $\mathbf{t}=[t_1,\dots,t_U]^\top \in {\mathbb{R}^+}^{U}$. This reformulation introduces the following $U$ additional constraints:
\begin{equation}\label{exp_const}
    t_u \leq \log_2(1+\gamma_u),~\forall u.
\end{equation}

Next, to decouple the interference and desired signal terms in $\gamma_u$, we define slack variables $\alpha_u,~\forall u,$ and $\beta_u,~\forall u,$ representing the SINR $\gamma_u$, and the total interference-plus-noise of UE~$u$, respectively. Accordingly, the following two new constraints will be introduced:
\begin{IEEEeqnarray*}{lcl}\label{betaalpha}
    \alpha_u \beta_u \leq \left|\sum_{b\in \mathcal{B}}{\left(\mathbf{h}^{b}_{u}\right)^H \mathbf{w}^{b}_{u}}\right|^2,~\forall u, \IEEEyesnumber \IEEEyessubnumber* 
    \label{eq:alphabetaileqPRi}\\
    \beta_u \geq \sum_{k \in \mathcal{U} \setminus \{u\}}\left|\sum_{b\in \mathcal{B}}{\left(\mathbf{h}^{b}_{u}\right)^H \mathbf{w}^{b}_{k}}\right|^2+\sigma^2_n,\,~\forall u.\label{eq:betaigeqIi}
\end{IEEEeqnarray*}

Observe that constraint \eqref{eq:betaigeqIi} is convex, whereas constraint \eqref{eq:alphabetaileqPRi} is nonconvex due to the presence of the norm function on its right-hand side and the product of two variables on the left-hand side. For now, we retain the nonconvex constraint \eqref{eq:alphabetaileqPRi} as it is, since the two-level distributed algorithm is capable of handling it.

To facilitate the distributed restructuring of the problem, we revisit the right-hand side of constraint \eqref{eq:betaigeqIi}. Specifically, the current form of \eqref{eq:betaigeqIi} begins with a summation over interfering UEs, whereas we prefer a formulation where the summation over BSs appears first to better align with the distributed structure. \textcolor{black}{To this end, we apply Jensen’s inequality, which follows from the convexity of the squared norm function, leading to the following upper bound for the interference term in the right-hand side of equation \eqref{eq:betaigeqIi}:}
\begin{equation}
    \sum_{k \in \mathcal{U} \setminus \{u\}}\left|\sum_{b\in \mathcal{B}}{\left(\mathbf{h}^{b}_{u}\right)^H \mathbf{w}^{b}_{k}}\right|^2 \leq \sum_{b\in \mathcal{B}}\sum_{k \in \mathcal{U} \setminus \{u\}}{\left|\left(\mathbf{h}^{b}_{u}\right)^H \mathbf{w}^{b}_{k}\right|^2},~\forall u.\label{Jensen's Inequality}
\end{equation}

Accordingly, constraint \eqref{eq:betaigeqIi} can be rewritten as
\begin{equation}
    \beta_u \geq \sum_{b\in \mathcal{B}}\sum_{k \in \mathcal{U} \setminus \{u\}}{\left|\left(\mathbf{h}^{b}_{u}\right)^H \mathbf{w}^{b}_{k}\right|^2}+\sigma^2_n,\,~\forall u.\label{eq:betaigeqIi_Jensen}
\end{equation}

To solve the problem through the two-level distributed algorithm, we introduce a few more new slack variables $\mathbf{s}_{b}=[s_{b,1},\dots,s_{b,U}]^\top \in {\mathbb{R}}^{U}~\text{and}~\mathbf{I}_{b}=[I_{b,1},\dots,I_{b,U}]^\top \in {\mathbb{R}^+}^{U}$, where $s_{b,u}$ and $I_{b,u}$ are defined as follows:
\begin{IEEEeqnarray*}{lcl}\label{S and I}
     s_{b,u} \leq {\left(\mathbf{h}^{b}_{u}\right)^H \mathbf{w}^{b}_{u}},~\forall u,~\forall b,\IEEEyesnumber \IEEEyessubnumber* \label{eq:PF_Const1_new1}\\
     I_{b,u} \geq \sum_{k \in \mathcal{U} \setminus \{u\}}{\left|\left(\mathbf{h}^{b}_{u}\right)^H \mathbf{w}^{b}_{k}\right|^2},~\forall u,~\forall b.\label{eq:PF_Const1_new2}
\end{IEEEeqnarray*}

Consequently, constraints \eqref{eq:alphabetaileqPRi} and \eqref{eq:betaigeqIi_Jensen} will equivalently be replaced as
\begin{IEEEeqnarray*}{lcl}\label{betaalpha_V2}
    \alpha_u \beta_u \leq \left|\sum_{b\in \mathcal{B}}s_{b,u}\right|^2,~\forall u, \IEEEyesnumber \IEEEyessubnumber* 
    \label{eq:alphabetaileqPRiV2}\\
    \beta_u \geq \sum_{b\in \mathcal{B}}{I_{b,u}}+\sigma^2_n,\,~\forall u. \label{eq:betaigeqIiV2}
\end{IEEEeqnarray*}

Furthermore, defining $\alpha_u$ as the lower bound for $\gamma_u$, constraints \eqref{eq:Original_Problem_const0} and \eqref{exp_const} will be replaced by \eqref{eq:MinSINR_V2} and \eqref{eq:exp_const_V2}, respectively.
\begin{IEEEeqnarray*}{lcl}\label{eq:PF_Const1_new}
     \alpha_u \geq \gamma_{\text{min}},~\forall u,\IEEEyesnumber \IEEEyessubnumber*\label{eq:MinSINR_V2} \\
     e^{t_u} \leq 1+\alpha_u,~\forall u.\label{eq:exp_const_V2}
\end{IEEEeqnarray*}

\textcolor{black}{It is important to note that although the reformulations are obtained via the upper bound in~\eqref{Jensen's Inequality}, all introduced auxiliary-variable inequalities \eqref{eq:alphabetaileqPRi} and \eqref{eq:betaigeqIi_Jensen}-\eqref{eq:PF_Const1_new} are active at the optimal solution. Furthermore, due to the mutual independence among channels of different BS, the Jensen-based approximation is tight on average, as discussed in Appendix~\ref{Appendix_A}. Therefore, the proposed reformulation provides a tight approximation of the original problem in terms of the optimal solution.}

Constraint \eqref{eq:exp_const_V2} is convex; however, to reduce computational complexity, it can be reformulated as a series of second-order cone (SOC) constraints using an exponential cone approximation, as shown in \eqref{Exp-cone}~\cite{Animesh-IEEEAccess}. \textcolor{black}{In this formulation, the parameter $\nu$ controls the accuracy of the approximation, and $k^u_i,\ i = 1, \dots, \nu + 4,\ \forall u$, are newly introduced slack variables. In particular, when $\nu = 4$, the SOC formulation achieves an accuracy exceeding $99.99\%$~\cite{nuImpact}.}
\begin{IEEEeqnarray*}{lcl}\label{Exp-cone}
    k^{u}_{\nu+4} \leq 1+\alpha_u,~\forall u,\IEEEyesnumber \IEEEyessubnumber* \label{Exp-cone1}\\
    \|[2+{t_u/2^{(\nu-1)}}~~1-k^{u}_1]^\top\|_2 \leq 1+k^{u}_1,~\forall u,\label{Exp-cone2}\\
    \|[5/3+{t_u/2^{\nu}}~~1-k^{u}_2]^\top\|_2 \leq 1+k^{u}_2,~\forall u,\label{Exp-cone3}\\
    \|[2k^{u}_1~~1-k^{u}_3]^\top\|_2 \leq 1+k^{u}_3,~\forall u,\label{Exp-cone4}\\
    19/72+k^{u}_2+1/24k^{u}_3 \leq k^{u}_4,~\forall u,\label{Exp-cone5}\\
    \|[2k^{u}_{i-1}~~1-k^{u}_i]^\top\|_2 \leq 1+k^{u}_i,~\forall u,~i=5,\dots,\nu+4.\qquad\label{Exp-cone6}
\end{IEEEeqnarray*}

\textcolor{black}{In this work, we do not employ the exponential cone approximation and instead solve the original constraint \eqref{eq:exp_const_V2} exactly. Therefore, the parameter $\nu$ has no impact on the results, and no approximation error is introduced.}
After aforementioned transformations, problem \eqref{eq:Original_Problem} can be reformulated as 
\begin{IEEEeqnarray*}{lcl}\label{eq:Relaxed Problem}
    &\underset{\mathbf{W},\mathbf{t},\boldsymbol{\beta},\boldsymbol{\alpha},\mathbf{S},\mathbf{I}}{\operatorname{maximize}}\,\, & \sum_{u\in\mathcal{U}}{\log(t_u)} \,  \IEEEyesnumber \IEEEyessubnumber* \label{eq:relaxed_Obj}\\
    &\text{s.t.} & \eqref{eq:Original_Problem_const1},\eqref{S and I},\eqref{betaalpha_V2}, \eqref{eq:PF_Const1_new},
\end{IEEEeqnarray*}
where $\boldsymbol{\beta}$, $\boldsymbol{\alpha}$, $\mathbf{S}$, and $\mathbf{I}$ are vectors and matrices that collect the variables $\beta_u$, $\alpha_u$, $s_{b,u}$, and $I_{b,u},~\forall u,~\forall b$, respectively. 
As can be observed, the set of constraints \eqref{eq:Original_Problem_const1} and \eqref{S and I} are defined per BS, while the set of constraints \eqref{betaalpha_V2} and \eqref{eq:PF_Const1_new} are defined per UE. However, these two sets of constraints are coupled through the variables $s_{b,u}$ and $I_{b,u}$. To address this coupling and facilitate the two-level distributed algorithm, we introduce new slack variables $\overline{\mathbf{s}}_{b} = [\overline{s}_{b,1}, \dots, \overline{s}_{b,U}]^\top \in {\mathbb{R}^+}^{U}$ and $\overline{\mathbf{I}}_{b} = [\overline{I}_{b,1}, \dots, \overline{I}_{b,U}]^\top \in {\mathbb{R}^+}^{U}$, which are associated with $\mathbf{s}_b$ and $\mathbf{I}_b$, respectively, such that they satisfy the following equality constraints:
\begin{IEEEeqnarray*}{lcl}\label{Equality Constraints 1}
     \mathbf{s}_b=\overline{\mathbf{s}}_b,~\forall b,\IEEEyesnumber \IEEEyessubnumber* \label{eq:s equality}\\
     \mathbf{I}_b=\overline{\mathbf{I}}_b,~\forall b.\label{eq:I equality}
\end{IEEEeqnarray*}

Consequently, we replace $s_{b,u}$ and $I_{b,u}$ in the per BS constraints \eqref{S and I} with $\overline{s}_{b,u}~\text{and}~\overline{I}_{b,u}$, as
\begin{IEEEeqnarray*}{lcl}\label{S and I V2}
     \overline{s}_{b,u} \leq \left(\mathbf{h}^{b}_{u}\right)^H \mathbf{w}^{b}_{u},~\forall u,~\forall b,\IEEEyesnumber \IEEEyessubnumber* \label{eq:PF_Const1_new1_bar}\\
     \overline{I}_{b,u} \geq \sum_{k \in \mathcal{U} \setminus \{u\}}{\left|\left(\mathbf{h}^{b}_{u}\right)^H \mathbf{w}^{b}_{k}\right|}^2,~\forall u,~\forall b.\label{eq:PF_Const1_new2_bar}
\end{IEEEeqnarray*}

Now, problem \eqref{eq:Relaxed Problem} can be reformulated as
\begin{IEEEeqnarray*}{lcl}\label{eq:Distributed_ReformulatedProblem}
    &\underset{\overline{\mathbf{S}},\overline{\mathbf{I}},\mathbf{W},\mathbf{t},\boldsymbol{\beta},\boldsymbol{\alpha},\mathbf{S},\mathbf{I}}{\operatorname{minimize}}\,\, & -\sum_{u\in\mathcal{U}}{\log(t_u)} \,  \IEEEyesnumber \IEEEyessubnumber* \label{eq:Relaxed_Objnew}\\
&\text{s.t.} & \eqref{betaalpha_V2},\eqref{eq:PF_Const1_new},\eqref{eq:Original_Problem_const1},\eqref{S and I V2},\eqref{Equality Constraints 1}.
\end{IEEEeqnarray*}

Problem \eqref{eq:Distributed_ReformulatedProblem} represents a distributed reformulation of the original problem \eqref{eq:Original_Problem}. In this reformulation, the decision variables and constraints are decomposed into local components, associated with individual BSs, and global components that coordinate their interactions. This separation enables each BS to solve its local subproblem independently, subject to a set of local constraints, while only a limited amount of information must be exchanged to satisfy the global constraints. Such a structure facilitates parallel decomposition and allows the problem to be solved efficiently in a distributed manner.
Accordingly, we define the compact variable sets as follows:
$\mathcal{X} = \{\mathbf{t}, \boldsymbol{\beta}, \boldsymbol{\alpha}, \mathbf{S}, \mathbf{I}\}$, representing the set of global optimization variables, and
$\overline{\mathcal{X}}=\{\overline{\mathbf{S}}, \overline{\mathbf{I}}, \mathbf{W}\}$, denoting the set of local optimization variables.
Accordingly, the associated constraint sets are defined as
\begin{IEEEeqnarray*}{lcl}\label{X and Xbar}
    \mathcal{C}:= \left\{ \boldsymbol{\nu} ~\middle|~ \text{\eqref{betaalpha_V2} and \eqref{eq:PF_Const1_new} are satisfied} \right\}
, \IEEEyesnumber \IEEEyessubnumber* \label{X set}\\
   \overline{\mathcal{C}} := \left\{ \boldsymbol{\nu} ~\middle|~ \text{\eqref{eq:Original_Problem_const1} and \eqref{S and I V2} are satisfied} \right\},\label{Xbar set}
\end{IEEEeqnarray*}
where $\mathcal{C}$ represents a nonconvex constraint set, and $\overline{\mathcal{C}}$ is a compact convex set. These sets are coupled through the equality constraints defined in \eqref{Equality Constraints 1}. The problems in the form of \eqref{eq:Distributed_ReformulatedProblem} can be expressed in the general distributed form as
\begin{IEEEeqnarray*}{lcl}\label{eq:General Distributed Problem}
    &\underset{\mathcal{X},\overline{\mathcal{X}}}{\operatorname{minimize}}\,\, & f(\mathcal{X})+g(\overline{\mathcal{X}}) \,  \IEEEyesnumber \IEEEyessubnumber* \label{eq:Original_obj}\\
    &\text{s.t.} & \mathbf{A}\mathcal{X}+\mathbf{B}\overline{\mathcal{X}}=0,\label{eq:Original_Equality}\\
    && \mathcal{X} \in \mathcal{C}.\label{eq:Original_Sets}
\end{IEEEeqnarray*}

It is worth noting that some common distributed algorithms, such as the two-block ADMM, can be employed to solve problems in the form of \eqref{eq:General Distributed Problem}. However, the convergence of the two-block ADMM algorithm relies on the following two conditions:
\begin{itemize}
\item Condition 1: $\operatorname{Im}(\mathbf{A}) \subseteq \operatorname{Im}(\mathbf{B})$, where $\operatorname{Im}(.)$ denotes the image space of the argument matrix.\footnote{The image space (or column space) of a matrix $\mathbf{A}$ is the set of all vectors that can be expressed as $\mathbf{A}\mathcal{X}$ for some vector $\mathcal{X}$.}
\item Condition 2: The last-block optimization variables are unconstrained and the objective function $g(\overline{\mathcal{X}})$ is Lipschitz differentiable.\footnote{A function $f$ is Lipschitz differentiable if it is differentiable and its gradient is Lipschitz continuous, meaning the gradient does not change too rapidly.}
\end{itemize}

In problem \eqref{eq:Distributed_ReformulatedProblem}, Condition 2 does not hold. Specifically, the last-block variable $\overline{\mathcal{X}}$ cannot be structured in a way that ensures Lipschitz differentiability. This difficulty arises because the variable set $\overline{\mathcal{X}}$ is constrained by a nonconvex set $\overline{\mathcal{C}}$. Even if the constraints defined by $\overline{\mathcal{C}}$ were relaxed and incorporated into the objective function through penalty terms, the resulting formulation would still yield a nonconvex and nonsmooth objective. \textcolor{black}{Consequently, the two-block ADMM approach, which has been widely used in the literature, does not guarantee convergence when applied to problem~\eqref{eq:Distributed_ReformulatedProblem}.} To overcome this limitation, the next section introduces the proposed distributed PFBWD algorithm, which adopts the two-level distributed framework developed in~\cite{TWoStepDistributed} to efficiently solve problem~\eqref{eq:Distributed_ReformulatedProblem}. This framework integrates a structured three-block ADMM method within an ALM-based outer loop, enabling scalable and efficient distributed optimization. Further details of the proposed algorithm are provided in the following section.
\section{Proposed Two-Level Distributed PFBWD Algorithm}\label{Sec:DistributedAlgorithm}
To develop the two-level distributed algorithm, we first reformulate problem \eqref{eq:Distributed_ReformulatedProblem} to make it suitable for the application of the three-block ADMM approach. Then, we design a two-level iterative algorithm, consisting of inner- and outer-level, to solve the reformulated problem. The inner-level algorithm utilizes the three-block ADMM method, with its iterations represented by $t$. Whereas the outer-level algorithm employs the ALM, with its iterations represented by $k$, to update the Lagrange multipliers and enforce the equality constraints. The aforementioned procedure is described in detail in the following subsections.
\subsection{Reformulation of Problem \eqref{eq:Distributed_ReformulatedProblem}}
For reformulation, we introduce auxiliary variables $\mathbf{z}_{l,b} \in \mathbb{R}^U,~l\in \mathcal{L}\triangleq\{s,I\},~\forall b,$ and replace the equality constraints in \eqref{Equality Constraints 1} with the following constraints as
\begin{IEEEeqnarray*}{lcl}\label{Equality Constraints 2}
    \mathbf{s}_b - \overline{\mathbf{s}}_b + \mathbf{z}_{s,b} = 0,\quad \forall b, \IEEEyesnumber \IEEEyessubnumber* \label{Equality 1}\\
    \mathbf{I}_b - \overline{\mathbf{I}}_b + \mathbf{z}_{I,b} = 0,\quad \forall b. \label{Equality 2}
\end{IEEEeqnarray*}

To ensure equivalence with the original problem \eqref{eq:Distributed_ReformulatedProblem}, we
Additionally, enforce the following constraints:
\begin{equation}
    \mathbf{z}_{l,b} = 0,~\forall l,~\forall b. \label{Z Constraints}
\end{equation}

The introduction of auxiliary variables, $\mathbf{z}_{l,b}$, yields two
important structural properties:
\begin{itemize}
    \item \textit{Property I:} The reformulated linear coupling constraints \eqref{Equality Constraints 2} now involve three blocks of variables. The third block corresponds to auxiliary variables associated with an identity matrix, whose image spans the full space. This structure satisfies Condition 1. 
    \item \textit{Property II:} If the constraints in \eqref{Z Constraints} are temporarily ignored, the problem becomes compatible with the standard three-block ADMM framework. In this case, the third block variable is unconstrained, thereby satisfying Condition 2 required for ADMM convergence.
\end{itemize}

To facilitate Property 2, we relax constraint \eqref{Z Constraints} by using ALM method, and thus, the objective function can now be written as
\begin{equation}\label{ALM Objective Function}
    f(\mathcal{X}, \mathcal{Z})\!=\!-\sum_{u\in\mathcal{U}} \log(t_u) + \sum_{b\in \mathcal{B}}\!\sum_{l\in\mathcal{L}}\!\left( \boldsymbol{\lambda}^\top_{l,b} \mathbf{z}_{l,b} + \frac{\rho_{o}}{2} \| \mathbf{z}_{l,b} \|^2 \right),
\end{equation}
where $\mathcal{Z}$ denotes collection of auxiliary variables $\mathbf{z}_{l,b}$, $\boldsymbol{\lambda}_{l,b}$ represents the Lagrange multipliers associated with the constraints $\mathbf{z}_{l,b} = 0$, and $\rho_{o} > 0$ is the penalty parameter for the outer-level ALM. Therefore, problem~\eqref{eq:Distributed_ReformulatedProblem}~is~reformulated~as
\begin{IEEEeqnarray*}{lcl}\label{eq:Relaxed Problem_V3}
    &\underset{\mathcal{X},\overline{\mathcal{X}},\mathcal{Z}}{\operatorname{minimize}} & ~f(\mathcal{X}, \mathcal{Z}) \IEEEyesnumber \IEEEyessubnumber* \label{eq:Relaxed_Obj} \\
    &\text{s.t.} & \mathcal{X} \in \mathcal{C}, \overline{\mathcal{X}} \in \overline{\mathcal{C}}, \text{\eqref{Equality Constraints 2}}.
\end{IEEEeqnarray*}

The augmented Lagrangian term in \eqref{ALM Objective Function} is both strongly convex in $\mathcal{Z}$ and Lipschitz differentiable, which simultaneously satisfies Conditions 1 and 2, and hence, ensures the convergence of the three-block ADMM algorithm to solve problem \eqref{eq:Relaxed Problem_V3}. However, the solution to the above problem does not necessarily ensure that constraint \eqref{Z Constraints} would be satisfied. To this end, we rely on ALM method which ensures variables $z_{l,b}$ to go to zero by updating $\lambda_{l,b}$. Specifically, this reformulation decomposes the original problem \eqref{eq:Distributed_ReformulatedProblem} into a two-level structure. The outer-level ALM algorithm iteratively drives the slack variables $\mathcal{Z}$ toward zero by updating the multipliers $\boldsymbol{\lambda}_{l,b}$ at each outer iteration indexed by $k$, thereby ensuring convergence to a stationary point of the original problem \eqref{eq:Distributed_ReformulatedProblem}~\cite{TWoStepDistributed}. Whereas, the inner-level algorithm solves the reformulated problem using the three-block distributed ADMM while ensuring convergence guarantees. The details of the inner-level algorithm are presented in the following subsection.

\begin{figure*}[t]
\setlength{\abovecaptionskip}{0pt}
\setlength{\belowcaptionskip}{0pt}
\centering
\parbox{0.95\textwidth}{
\begin{equation}
L(\mathcal{X},\overline{\mathcal{X}},\mathcal{Z},\boldsymbol{\Psi}) = f(\mathcal{X},\mathcal{Z})+\sum_{b\in\mathcal{B}}\sum_{l\in\mathcal{L}}\left({\boldsymbol{\psi}^\top_{l,b}}\mathbf{r}_{l,b}+\frac{\rho_{l,b}}{2} \|\mathbf{r}_{l,b}\|^2 \right)=-\sum_{u\in\mathcal{U}}{\log(t_{u})}+\sum_{b\in\mathcal{B}}{L^b(\mathcal{X}_b,\overline{\mathcal{X}}_b,\mathbf{Z}_b,\boldsymbol{\psi}_b)}, \label{eq:augmented_lagrangian1}
\end{equation}
\begin{IEEEeqnarray}{rCl}
\text{where:} && \nonumber \\
L^b(\mathcal{X}_b,\overline{\mathcal{X}}_b,\mathbf{Z}_b,\boldsymbol{\psi}_b) &\triangleq& \sum_{l\in\mathcal{L}}\left(\boldsymbol{\lambda}^\top_{l,b} \mathbf{z}_{l,b} +\frac{\rho_{o}}{2}\|\mathbf{z}_{l,b}\|^2 + {\boldsymbol{\psi}^\top_{l,b}}\mathbf{r}_{l,b}+ \frac{\rho_{l,b}}{2} \|\mathbf{r}_{l,b}\|^2\right),~\forall b . \label{eq:augmented_lagrangian_per base station}
\end{IEEEeqnarray}
\vspace{0.5em}
\hrule
}
\end{figure*}
\subsection{Inner-Level Three-Block ADMM Algorithm}
Given $\boldsymbol{\lambda}^{(k)}_{l,b}$ and $\rho^{(k)}_{o}$ as the Lagrange multipliers and penalty parameter for the $k$th outer-level iteration, the augmented Lagrangian function corresponding to the $k$th inner-level problem is defined in \eqref{eq:augmented_lagrangian1}. In \eqref{eq:augmented_lagrangian1} and \eqref{eq:augmented_lagrangian_per base station}, matrices $\mathbf{Z}_b,~\forall b,$ include the vectors $\mathbf{z}_{l,b},~\forall l$, and $\boldsymbol{\psi}_{l,b}$ denotes the Lagrange multipliers associated with the equality constraints in \eqref{Equality Constraints 2}, which are reflected by the defined residuals $\mathbf{r}_{l,b}$, as
\begin{IEEEeqnarray*}{lcl}\label{Residuals}
\mathbf{r}_{s,b} = \mathbf{s}_b - \overline{\mathbf{s}}_b + \mathbf{z}_{s,b},~\forall b, \IEEEyesnumber \IEEEyessubnumber* \label{residual_s}\\
\mathbf{r}_{I,b} = \mathbf{I}_b - \overline{\mathbf{I}}_b + \mathbf{z}_{I,b},~\forall b. \label{residual_I}
\end{IEEEeqnarray*}
 
According to the three-block ADMM approach, problem \eqref{eq:Relaxed Problem_V3} can be partitioned into three variable blocks and solved iteratively in an alternative manner until the convergence criteria are met. The three blocks, and the Lagrange multipliers are updated according to the following equations:
\begin{IEEEeqnarray}{l}
\IEEEyesnumber
\IEEEyessubnumber*
\text{ADMM Block 1:} \nonumber\\*
\quad \underset{\mathcal{X}\in \mathcal{C}}{\operatorname{minimize}} \,~L(\mathcal{X}, \overline{\mathcal{X}}^{(t-1)}, \mathcal{Z}^{(t-1)}, \boldsymbol{\Psi}^{(t-1)}) \label{eq:Block 1} \\[0.5em]
\text{ADMM Block 2:}\nonumber \\*
\quad \underset{\overline{\mathcal{X}}\in \overline{\mathcal{C}}}{\operatorname{minimize}} \,~ L(\mathcal{X}^{(t)}, \overline{\mathcal{X}}, \mathcal{Z}^{(t-1)}, \boldsymbol{\Psi}^{(t-1)}) \label{eq:Block 2}\\[0.5em]
\text{ADMM Block 3:} \nonumber\\*
\quad \underset{\mathcal{Z}}{\operatorname{minimize}} \,~L(\mathcal{X}^{(t)}, \overline{\mathcal{X}}^{(t)}, \mathcal{Z}, \boldsymbol{\Psi}^{(t-1)}) \label{eq:Block 3}\\[0.5em]
\text{Lagrange Multiplier Update:} \nonumber\\*
\quad \boldsymbol{\psi}^{(t)}_{l,b} \leftarrow \boldsymbol{\psi}^{(t-1)}_{l,b} + \rho^{(k)}_{l,b} \mathbf{r}^{(t)}_{l,b},~\forall l,~\forall b.\label{eq:LagrangeUpdate}
\end{IEEEeqnarray}

In the remainder of this section, we provide a detailed explanation of the optimization problem for each block, followed by the pseudocode of the entire inner-level procedure in Algorithm~\ref{alg:Inner ADMM_Proposed}.
\subsubsection{Inner-level ADMM block~1 optimization problem}
The optimization problem solved in the $t$th iteration of the inner-level algorithm block 1 is given by \eqref{eq:Block 1}. In problem \eqref{eq:Block 1}, the constraint set $\mathcal{C}$ includes the nonconvex constraint \eqref{eq:alphabetaileqPRiV2}. Consequently, problem \eqref{eq:Block 1} cannot be solved directly by standard solvers and requires either convex approximations or iterative resolution via SCA. However, since the problem size is small, the iterative algorithm is very fast. To facilitate this, we introduce new slack variables $\mathbf{p} = [p_1, \dots, p_U]^\top \in \mathbb{R}^U$ and $\mathbf{q} = [q_1, \dots, q_U]^\top \in \mathbb{R}^U$, representing the real and imaginary parts, respectively, of the summation on the right-hand side of \eqref{eq:alphabetaileqPRiV2}, defined as
\begin{equation}\label{P and q}
     p_{u} = \Re{\left(\sum_{b\in \mathcal{B}}s_{b,u}\right)},~q_{u} = \Im{\left(\sum_{b\in \mathcal{B}}s_{b,u}\right)},~\forall u.
\end{equation}

In this way, the right-hand side of inequality \eqref{eq:alphabetaileqPRiV2} will be replaced with $p_u^2+q_u^2$. Accordingly, moving $\beta_u$ to the right-hand side and replacing the right-hand side with its approximated first-order Taylor series, the constraint can be approximated as
\begin{equation}{\label{first Taylor}}
     \alpha_u \leq \frac{2{p^{(m)}_{u}}}{{\beta^{(m)}_{u}}}\left(p_{u}-{p^{(m)}_{u}}\right)+\frac{2{q^{(m)}_{u}}}{{\beta^{(m)}_u}}\left(q_{u}-{q^{(m)}_{u}}\right)+\frac{\left({p^{(m)}_{u}}\right)^2+\left({q^{(m)}_{u}}\right)^2}{\beta^{(m)}_u}\left(1-\frac{\beta_u-{\beta^{(m)}_u}}{{\beta^{(m)}_u}}\right),~\forall u,
\end{equation}
where the superscript $(m)$ refers to the value of the corresponding variable in the $m$th iteration of the SCA process. In this way, the $m$th SCA iteration approximate problem for problem \eqref{eq:Block 1} will be as
\begin{IEEEeqnarray*}{lcl}\label{eq:Block 1_V2}
    &\underset{\mathcal{X}}{\operatorname{minimize}} \,\, & L(\mathcal{X},\overline{\mathcal{X}}^{(t-1)},\mathcal{Z}^{(t-1)},\boldsymbol{\Psi}^{(t-1)})\,  \IEEEyesnumber \IEEEyessubnumber* \label{eq:Block 1_Obj_V2}\\
    &\text{s.t.} & \eqref{eq:betaigeqIiV2},\eqref{eq:PF_Const1_new},\eqref{P and q},\eqref{first Taylor}.
\end{IEEEeqnarray*}

Algorithm~\ref{alg:ADMM Block 1 SCA} describes the first-block SCA-based algorithm. At each inner-level ADMM iteration $t$, Algorithm~\ref{alg:ADMM Block 1 SCA} solves the approximated problem \eqref{eq:Block 1_V2} iteratively for the optimal solution. The optimal point of the SCA-based approximated problem satisfies the Karush-Kuhn-Tucker (KKT) conditions of the original problem \eqref{eq:Block 1}~[\cite{SCA} Proposition 3.2]. Therefore, the solution of Algorithm~\ref{alg:ADMM Block 1 SCA} is a stationary point of the problem \eqref{eq:Block 1}, and hence a feasible solution. The SCA-based Algorithm~\ref{alg:ADMM Block 1 SCA} stops upon convergence, which is defined as when a tolerable change in the objective function value is achieved (less than $\epsilon$), or the maximum number of SCA iterations, $N_\text{iter}$, is reached, whichever first. This algorithm can be solved centrally or in a distributed manner at each BS using the same initial and broadcasted values.

\begin{algorithm}[h!]
\caption{\small SCA-based algorithm for ADMM~Block~1 iteration $t$.}\label{alg:ADMM Block 1 SCA}
    \begin{algorithmic}[1]
    \STATE Initialize $\mathbf{p}^{(0)}$, $\mathbf{q}^{(0)}$, $\boldsymbol{\beta}^{(0)},$ and $m \leftarrow 0$.
    \REPEAT
        \STATE Solve (\ref{eq:Block 1_V2}) to find $\mathcal{X}^{(t)*}$.
        \STATE Update $\mathbf{p}^{(m+1)} \leftarrow \mathbf{p}^{(m)*}$, $\mathbf{q}^{(m+1)} \leftarrow \mathbf{q}^{(m)*}$,       $\boldsymbol{\beta}^{(m+1)} \leftarrow \boldsymbol{\beta}^{(m)^*},$~$m \leftarrow m+1$.
    \UNTIL{convergence or $m > N_{\text{iter}}$.}
    \end{algorithmic}
\end{algorithm}
\color{black}
\subsubsection{Inner-level ADMM block 2 optimization problem}
The optimization problem to be solved in ADMM Block 2 is given in \eqref{eq:Block 2}. It is a convex problem that can be solved using standard solvers. Moreover, the structure of ADMM Block 2 lends itself well to decomposition into $(B+1)$ subproblems, each of which can be solved locally by a BS in parallel. In particular, the constraint set $\overline{\mathcal{C}}$ imposes per BS constraints, and the objective function $L(\mathcal{X}^{(t)},\overline{\mathcal{X}},\mathcal{Z}^{(t-1)},\boldsymbol{\Psi}^{(t-1)})$ is the sum of independent per–BS terms. Therefore, we can separate the objective and constraints so that each BS solves its subproblem in parallel. In this way, the optimization problem at each BS $b$ will be as
\begin{IEEEeqnarray*}{lcl}\label{eq:Block 2 per base station}
    &\underset{\overline{\mathcal{X}}_b}{\operatorname{minimize}} \,\, & L^b(\mathcal{X}^{(t)}_b,\overline{\mathcal{X}}_b,\mathbf{Z}^{(t-1)}_b,\boldsymbol{\psi}^{(t-1)}_b)\,  \IEEEyesnumber \IEEEyessubnumber* \label{eq:Block 2_Obj}\\
    &\text{s.t.} & {\|\mathbf{W}^b \|^2_F}\leq P^{\text{max}}_b,\\
    && \overline{s}_{b,u} \leq \left(\mathbf{h}^{b}_{u}\right)^H \mathbf{w}^{b}_{u},~\forall u,\\
    && \overline{I}_{b,u} \geq \sum_{k \in \mathcal{U} \setminus \{u\}}{\left|\left(\mathbf{h}^{b}_{u}\right)^H \mathbf{w}^{b}_{k}\right|}^2,~\forall u,
\end{IEEEeqnarray*}
where $L^b(\mathcal{X}^{(t)}_b,\overline{\mathcal{X}}_b,\mathbf{Z}^{(t-1)}_b,\boldsymbol{\psi}^{(t-1)}_b)$ is defined according to \eqref{eq:augmented_lagrangian_per base station}.

\subsubsection{Inner-level ADMM block 3 optimization problem}
The third block of the Inner-level ADMM algorithm, at iteration $t$, solves the problem \eqref{eq:Block 3} to obtain the $\mathcal{Z}^{(t)}$ values. Problem \eqref{eq:Block 3} represents a convex optimization problem, and with the same logic as block 2, can be decomposed to $(B+1)$ independent optimization problems which can be solved in parallel by each BS $b$. The ADMM block 3 optimization problem to be solved at each BS $b$ will be as
\begin{IEEEeqnarray*}{lcl}\label{eq:Block 3 per base station}
    &\underset{\mathbf{Z}_b}{\operatorname{minimize}} \,\, & L^b(\mathcal{X}^{(t)}_b, \overline{\mathcal{X}}^{(t)}_b, \mathbf{Z}_b, \boldsymbol{\Psi}^{(t-1)}_b)\,  \IEEEyesnumber \IEEEyessubnumber* \label{eq:Block 3_Obj per base}
\end{IEEEeqnarray*}

\begin{figure*}[t]
\begin{IEEEeqnarray*}{lCl}\label{eq:Inner_Stopping_Criteria}
\left\|\sum_{b\in \mathcal{B}} \Bigl[
\rho^{(k)}_{s,b} \bigl( -\overline{\mathbf{s}}^{(t)}_b + \mathbf{z}^{(t)}_{s,b} 
+ \overline{\mathbf{s}}^{(t-1)}_{b} - \mathbf{z}^{(t-1)}_{s,b} \bigr) + \rho^{(k)}_{I,b} \bigl( -\overline{\mathbf{I}}^{(t)}_b + \mathbf{z}^{(t)}_{I,b} 
+ \overline{\mathbf{I}}^{(t-1)}_{b} - \mathbf{z}^{(t-1)}_{I,b} \bigr)
\Bigr]\right\| \leq \epsilon_1, \qquad\IEEEyesnumber \IEEEyessubnumber* \\
\left\|\sum_{b\in \mathcal{B}} \sum_{l\in \mathcal{L}}\Bigl[
\rho^{(k)}_{l,b} \bigl(\mathbf{z}^{(t-1)}_{l,b} - \mathbf{z}^{(t)}_{l,b}\bigr)\Bigr]\right\|\leq \epsilon_{2},\qquad\\
\left\|\mathbf{r}^{(t)}_{l,b}\right\| \leq \epsilon_{l,b},~\forall l,~\forall b.
\end{IEEEeqnarray*}
\hrule
\end{figure*}
\begin{algorithm}[t!]
\caption{The $k$th inner-level ADMM Algorithm.}
\label{alg:Inner ADMM_Proposed}
\begin{algorithmic}[1]
    \STATE \textbf{Input:} $\mathbf{I}^{(k-1)}$, $\mathbf{S}^{(k-1)}$, $\mathcal{Z}^{(k-1)}$, $\boldsymbol{\lambda}_{l,b}^{(k)},~\rho_{o}^{(k)}$.
    \STATE \textbf{Output:} $\{\mathbf{W}^{b*}\}_{b=1}^{B+1}$, $\mathbf{I}^{(k)}$, $\mathbf{S}^{(k)}$, and $\mathcal{Z}^{(k)}$.
    
    \STATE \textbf{Initialization:}
    \STATE \hspace{-1.4em} \null \hspace{2em}$t \leftarrow 1$; $n \leftarrow 1$, $\rho^{(k)}_{l,b}=\delta \rho^{(k)}_{{o}}$.
    \STATE \hspace{-1.4em} \null \hspace{2em}$\mathbf{I}^{(0)} \leftarrow \mathbf{I}^{(k-1)}$, $\mathbf{S}^{(0)} \leftarrow \mathbf{S}^{(k-1)}$, $\mathcal{Z}^{(0)} \leftarrow \mathcal{Z}^{(k-1)}$.
    \STATE \hspace{-1.4em} \null \hspace{2em}Choose $\boldsymbol{\psi}_{l,b}^{(0)},~\forall b,$ such that:
    \STATE \hspace{-1.4em} \null \hspace{4em}$\boldsymbol{\lambda}_{l,b}^{(k)} + \rho_{o}^{(k)} \mathbf{z}_{l,b}^{(0)} + \boldsymbol{\psi}_{l,b}^{(0)} = 0$ for $l \in \mathcal{L},~\forall b$.
    \STATE \hspace{-1.4em} \null \hspace{2em}Set tolerance thresholds $\{\epsilon_1,\epsilon_2,\epsilon_{l,b}\}$.
    \WHILE{stopping criteria \eqref{eq:Inner_Stopping_Criteria} is not met}
        \FORALL{BSs $b \in \mathcal{B}$ \textbf{(in parallel)}}
            \STATE ADMM Block 1:\\
            \hspace{1em} Solve the problem \eqref{eq:Block 1_V2} using Algorithm~\ref{alg:ADMM Block 1 SCA}.
            \STATE ADMM Block 2:\\
            \hspace{1em} Solve problem~\eqref{eq:Block 2 per base station}.    
            \STATE ADMM Block 3:\\
            \hspace{1em} Solve problem \eqref{eq:Block 3 per base station}.
            \STATE Lagrange Multiplier Update:
            \STATE \hspace{1em}$\boldsymbol{\psi}^{(t)}_{l,b} \leftarrow \boldsymbol{\psi}^{(t-1)}_{l,b} + \rho^{(k)}_{l,b} \mathbf{r}^{(t)}_{l,b},~\forall l.$
            \STATE Broadcast $\mathbf{s}_b^{(t)}$, $\mathbf{I}_b^{(t)}$, and  $\mathbf{Z}_{b}^{(t)},$ to all BSs.
        \ENDFOR
        \STATE $t \leftarrow t + 1$
    \ENDWHILE
    % \STATE Return $\mathbf{I}^{(t)}$, $\mathbf{S}^{(t)}$, $\mathcal{Z}^{(t)}$.
\end{algorithmic}
\end{algorithm}

Algorithm~\ref{alg:Inner ADMM_Proposed} summarizes the $k$th inner-level three-block ADMM procedure for solving the distributed PFBWD problem. At each outer-level iteration indexed by $k$, the algorithm initializes local variables and Lagrange multipliers, then alternatively updates the three variable blocks, $\mathcal{X}$, $\overline{\mathcal{X}}$, and $\mathcal{Z}$, while enforcing the equality constraints via augmented Lagrangian terms. This structure enables the BSs to solve their local subproblems in parallel during Blocks 2 and 3. This is further explained in detail later in this section. Accordingly, the Lagrange multipliers are updated using the residuals of the coupling constraints, and only a small amount of information is exchanged among the BSs. The procedure repeats until the specified stopping criteria are satisfied, at which point the updated variables are returned for use in the next outer-level ALM iteration. According to \cite{TWoStepDistributed}, Algorithm~\ref{alg:Inner ADMM_Proposed} terminates once it identifies a point $(\mathcal{X}^{(t)}, \overline{\mathcal{X}}^{(t)}, \mathcal{Z}^{(t)})$ that satisfies the stopping conditions specified in \eqref{eq:Inner_Stopping_Criteria}.
The rationale behind the stopping criteria \eqref{eq:Inner_Stopping_Criteria} will be discussed in Section \ref{Sec:Complexity}.

\subsection{Proposed Distributed Two-Level PFBWD Algorithm}
In this section, we integrate the previously described components to present the proposed distributed two-level \mbox{PFBWD} algorithm. As shown in Algorithm~\ref{alg:alg_Distributed_Proposed}, the algorithm is designed for a HAPS-empowered vHetNet consisting of $B$ MBSs and one HAPS. It adopts an outer–inner structure, where the outer-level updates the Lagrange multipliers and penalty parameters, while the inner-level, executed via the three-block ADMM in Algorithm~\ref{alg:Inner ADMM_Proposed}, handles the main optimization.
At each outer iteration $k$, the central controller broadcasts initial variables to all BSs. Each BS then solves its local subproblem in parallel to update local variables and interference estimates. The central controller collects these updates to adjust the Lagrange multipliers and the penalty parameter based on the residual norm, thereby enhancing convergence.
The proposed approach is both efficient and scalable, as it exploits problem separability across BSs to enable parallel computation without requiring CSI exchange with the central node. This significantly reduces signaling overhead, which is especially beneficial in networks with large-scale antenna arrays, particularly at the HAPS. The stopping criteria for both inner- and outer-levels, along with a complexity analysis, are provided in Section~\ref{Sec:Complexity}.
\begin{algorithm}[h!]
\caption{Proposed Two-Level Distributed PFBWD Algorithm for a HAPS-Empowered vHetNet with $B$ MBSs and One HAPS.}
\label{alg:alg_Distributed_Proposed}
\begin{algorithmic}[1]
    \STATE \textbf{Input:} $U$,~$B$,~$\sigma^2_n$.
    \STATE \textbf{Output:} $\{\mathbf{W}^{b*}\}_{b=1}^{B+1}$
    \STATE Initialize:
    \STATE \hspace{1em} $\mathbf{S}^{(0)},~\mathcal{Z}^{(0)},~\boldsymbol{\lambda}_{l,b}^{(0)} \in \mathbb{R}^{(B+1) \times U},~\forall l,~\forall b$,
    \STATE \hspace{1em} $\mathbf{I}^{(0)} \in \mathbb{R}_+^{(B+1) \times U}$,~$\rho^{(1)}_{o} > 0,~\omega \in [0,1),~\gamma > 1,~k \leftarrow 1$
    
    \WHILE{stopping criterion is not met}
        \STATE Broadcast current variables to all BSs.
        \STATE Use $(\mathbf{S}^{(k-1)}, \mathbf{I}^{(k-1)}, \mathcal{Z}^{(k-1)})$ as the initial point and solve Algorithm~\ref{alg:Inner ADMM_Proposed}.
        \STATE Obtain updated values $(\mathbf{S}^{(k)}, \mathbf{I}^{(k)}, \mathcal{Z}^{(k)})$.
        \FORALL{BSs $b \in \mathcal{B}$ \textbf{(in parallel)}}
            \FORALL{$l \in \mathcal{L}$}
                \STATE $\boldsymbol{\lambda}_{l,b}^{(k+1)} \leftarrow \text{Proj} \bigl(\boldsymbol{\lambda}_{l,b}^{(k)} + \rho^{(k)}_{o} \mathbf{z}_{l,b}^{(k)}\bigr)$
            \ENDFOR
        \ENDFOR
        \IF{$\|\mathbf{Z}_b^{(k)}\| \geq \omega \|\mathbf{Z}_b^{(k-1)}\|,~\forall b,$}
            \STATE $\rho^{(k+1)}_{o} \leftarrow \gamma \rho^{(k)}_{o}$
        \ENDIF
        \STATE $k \leftarrow k + 1$
    \ENDWHILE
\end{algorithmic}
\end{algorithm}
\section{Convergence and Computational Complexity Analysis of the Proposed PFBWD Algorithm}\label{Sec:Complexity}
In this section, we will provide details regarding the convergence and computational complexity analysis of the proposed distributed PFBWD algorithm.
\vspace{-2mm}
\subsection{Convergence}
To establish the convergence of the proposed two-level distributed PFBWD algorithm, we verify that the reformulated problem~\eqref{eq:Relaxed Problem_V3} satisfies the conditions required for the convergence results in~\cite{TWoStepDistributed}. Specifically, we show that the three assumptions establishing the convergence of the two-level distributed algorithm are satisfied for the problem in this work, i.e., problem~\eqref{eq:Relaxed Problem_V3}.

\color{black}
\begin{assumption}
The objective function $f(\mathcal{X})$ is continuously differentiable, $\mathcal{C}$ is a compact set, and $\overline{\mathcal{C}}$ is convex and compact.

\begin{proof} The objective function $f(\mathcal{X}) = -\sum_{u \in \mathcal{U}} \log(t_u)$
is continuously differentiable for $t_u > 0,~\forall u$. Since the constraints of the problem explicitly enforce $t_u>0,~\forall u$, $f(\mathcal{X})$ is continuously differentiable over the entire feasible set. 
Next, consider the constraint sets $\mathcal{C}$ and $\overline{\mathcal{C}}$, defined in~\eqref{X and Xbar}. The set $\overline{\mathcal{C}}$ is convex because it is described by SOC constraints in~\eqref{eq:Original_Problem_const1} and~\eqref{S and I V2}, which define convex feasible regions. Moreover, these constraints impose explicit upper and lower bounds on the optimization variables $\bar{\mathcal{X}}$, implying that $\overline{\mathcal{C}}$ is closed and bounded, and hence compact.
Similarly, the set $\mathcal{C}$ is compact since the constraints~\eqref{betaalpha_V2} and~\eqref{eq:PF_Const1_new} impose finite upper and lower bounds on all global optimization variables $\mathcal{X}$. As a result, $\mathcal{C}$ is closed and bounded, and therefore compact.
\end{proof}
\end{assumption}

\begin{assumption}
Problem~\eqref{eq:Relaxed Problem_V3} is feasible, and the set of stationary points of problem \eqref{eq:Distributed_ReformulatedProblem} is nonempty.

\begin{proof} Under Assumption~1, the constraint sets $\mathcal{C}$ and $\overline{\mathcal{C}}$ are compact and nonempty. Since problem~\eqref{eq:Relaxed Problem_V3} only enforces the membership constraints $\mathcal{X}\in\mathcal{C}$ and $\overline{\mathcal{X}}\in\overline{\mathcal{C}}$, together with the linear coupling constraints~\eqref{Equality Constraints 2}, feasibility reduces to the existence of at least one point satisfying these sets simultaneously. Because the original problem \eqref{eq:Distributed_ReformulatedProblem} admits a feasible solution, there exists at least one pair $(\mathcal{X},\overline{\mathcal{X}})$ satisfying the defining constraints of $\mathcal{C}$ and $\overline{\mathcal{C}}$. Moreover, the linear constraints~\eqref{Equality Constraints 2} do not restrict feasibility, as the auxiliary variables $\mathbf{z}_{l,b},~\forall l,~\forall b,$ are unconstrained in problem~\eqref{eq:Relaxed Problem_V3} due to the relaxation of~\eqref{Z Constraints} via the augmented Lagrangian formulation. Therefore, the feasible set of problem~\eqref{eq:Relaxed Problem_V3} is nonempty, and problem~\eqref{eq:Relaxed Problem_V3} is feasible.

Under Assumption~1, the feasible set of problem~\eqref{eq:Distributed_ReformulatedProblem} is nonempty, compact, and closed. Moreover, the objective function $f(\mathcal{X})$ is continuous and continuously differentiable over the feasible region. By the Weierstrass extreme value theorem, problem~\eqref{eq:Distributed_ReformulatedProblem} admits at least one global minimizer over the feasible region. Any global minimizer is, in particular, a local minimizer. Since all constraints of problem~\eqref{eq:Distributed_ReformulatedProblem} are continuously differentiable and standard constraint qualifications hold for the feasible region defined by $\mathcal{C}$ and $\overline{\mathcal{C}}$, the KKT conditions are necessary optimality conditions at any local minimizer.
Therefore, there exists at least one feasible point satisfying the KKT conditions of problem~\eqref{eq:Distributed_ReformulatedProblem}. By definition, such a point is a stationary point. Hence, the set of stationary points of problem~\eqref{eq:Distributed_ReformulatedProblem} is nonempty.
\end{proof}

\end{assumption}

\color{black}
\begin{assumption}
Given $\boldsymbol{\lambda}^{(k)}_{l,b}$, $\rho^{(k)}_{o}$, and $\rho^{(k)}_{l,b},~\forall l,~\forall b$, the first block Algorithm~\ref{alg:ADMM Block 1 SCA} can find a stationary solution $\mathcal{X}^{(t)}$ such that 
\begin{equation}
0 \in \partial_x L(\mathcal{X}^{(t)}, \overline{\mathcal{X}}^{(t-1)}, \mathcal{Z}^{(t-1)}, \boldsymbol{\Psi}^{(t-1)}),\nonumber
\end{equation}
and
\begin{equation}
L(\mathcal{X}^{(t)}, \overline{\mathcal{X}}^{(t-1)}, \mathcal{Z}^{(t-1)}, \boldsymbol{\Psi}^{(t-1)}) \leq L(\mathcal{X}^{(t-1)}, \overline{\mathcal{X}}^{(t-1)}, \mathcal{Z}^{(t-1)}, \boldsymbol{\Psi}^{(t-1)}) < +\infty, \nonumber
\end{equation}
for all $t \in \mathbb{Z}_{++}$.

\begin{proof}At inner-level iteration $t$, the inner-level ADMM block 1 Algorithm~\ref{alg:ADMM Block 1 SCA}
solves a nonconvex problem through SCA.
By construction, the solution of Algorithm~\ref{alg:ADMM Block 1 SCA} is obtained by solving a convex surrogate problem generated via SCA. Each surrogate problem is a tight local approximation of the original nonconvex subproblem~\eqref{eq:Block 1} and shares the same first-order behavior at the previous iterate $\mathcal{X}^{(t-1)}$.
Since the surrogate problem is convex and feasible, its optimal solution $\mathcal{X}^{(t)}$ satisfies the first-order optimality conditions. Moreover, due to the consistency and majorization properties
of the SCA framework~\cite{SCA}, satisfaction of the surrogate optimality conditions implies stationarity of $\mathcal{X}^{(t)}$ with respect to the original augmented Lagrangian, i.e.,
\[
0 \in \partial_x L(\mathcal{X}^{(t)}, \overline{\mathcal{X}}^{(t-1)},
\mathcal{Z}^{(t-1)}, \boldsymbol{\Psi}^{(t-1)}).
\]

Finally, since the augmented Lagrangian is bounded from below over the compact feasible set (following Assumption~1), the sequence of objective values is finite. Therefore, at each iteration $t$, the first-block update produces a stationary solution with a non-increasing augmented Lagrangian value.

\end{proof}
\end{assumption}

\color{black}
\subsubsection{Convergence of the inner-level algorithm~\ref{alg:Inner ADMM_Proposed}} 
According to [\cite{TWoStepDistributed},~Proposition 1], under the above assumptions, the inner-level Algorithm~\ref{alg:Inner ADMM_Proposed} terminates at a point $(\mathcal{X}^{(k)},\overline{\mathcal{X}}^{(k)},\mathcal{Z}^{(k)})$ satisfying the conditions \eqref{Stationary Point Conditions}. 
\subsubsection{Convergence of the outer-level algorithm~\ref{alg:alg_Distributed_Proposed}} Under assumptions 1-3, according to [\cite{TWoStepDistributed},~Theorem 1], the outer-level Algorithm has a limit point $(\mathcal{X}^{*},\overline{\mathcal{X}}^*,\mathcal{Z}^*)$ which is either feasible to the original problem, i.e., $\mathcal{Z}^*=0,$ or $(\mathcal{X}^{*},\overline{\mathcal{X}}^*)$ is a stationary point of the problem
\begin{equation}
    \underset{\mathcal{X}\in \mathcal{C},~\overline{\mathcal{X}}\in \overline{\mathcal{C}}}{\operatorname{minimize}} \, \,~ \frac{1}{2}\left\|\mathcal{X}-\overline{\mathcal{X}}\right\|^2.
\end{equation}

According to [\cite{TWoStepDistributed},~Theorem 2], if $\boldsymbol{\Psi}^*$ is the limit Lagrange multiplier, associated with the point $(\mathcal{X}^{*},\overline{\mathcal{X}}^*,\mathcal{Z}^*)$, the point $(\mathcal{X}^{*},\overline{\mathcal{X}}^*,\boldsymbol{\Psi}^*)$ is a stationary point of the original problem \eqref{eq:Original_Problem}.

Accordingly, the stopping criteria for the inner-level and outer-level algorithms are presented as follows.
% \subsection{Stopping Criteria}
\setcounter{subsubsection}{0}
\subsubsection{Stopping Criteria for the inner-level algorithm \ref{alg:Inner ADMM_Proposed}}
The $k$th inner-level algorithm aims to find an approximate stationary solution $(\mathcal{X}^{(k)},\overline{\mathcal{X}}^{(k)},\mathcal{Z}^{(k)})$ of the problem \eqref{eq:Relaxed Problem_V3}. Therefore, defining $f(\mathcal{X})=-\sum_{u\in\mathcal{U}}\log(t_u)$, according to the first order optimality condition of constrained optimization problems \cite{BoydBook}, at convergence of the $t$th inner-level ADMM problem, the generated point $(\mathcal{X}^{(k)},\overline{\mathcal{X}}^{(k)},\mathcal{Z}^{(k)})$ should satisfy the first order optimality condition of the problem \eqref{eq:Relaxed Problem_V3} in the sense that there exist $d^{(k)}_j,~\forall j \in \{1,2\},~d^{(k)}_{3,b},~\text{and}~d^{(k)}_{l,b}$, such that
\begin{IEEEeqnarray*}{lCl}\label{Stationary Point Conditions}
d^{(k)}_1 \in \nabla f(\mathcal{X}^{(k)}) +  \sum_{b\in \mathcal{B}} \sum_{l\in\mathcal{L}}\boldsymbol{\psi}^{(k)}_{l,b}  + N_{\mathcal{C}}(\mathcal{X}^{(k)}), \qquad \IEEEyesnumber \IEEEyessubnumber* \\
d^{(k)}_2 \in  -\sum_{b\in \mathcal{B}} \sum_{l\in\mathcal{L}}\boldsymbol{\psi}^{(k)}_{l,b} + N_{\overline{\mathcal{C}}}(\overline{\mathcal{X}}^{(k)}),\qquad \\ 
0 =\mathbf{\lambda}^{(k)}_{l,b} + \rho^{(k)}_{{o}} \mathbf{z}^{(k)}_{l,b} + \boldsymbol{\psi}^{(k)}_{l,b} ,~\forall b,~\forall l,\qquad\\
d^{(k)}_{l,b} = \mathbf{r}^{(k)}_{l,b},~\forall l,~\forall b, \qquad\\
\| d^{(k)}_{j} \| \leq \epsilon^{(k)}_{j},~\forall j \in \{1,2\},~\| d^{(k)}_{l,b} \| \leq \epsilon^{(k)}_{l,b},~\forall l,~\forall b,\qquad
\end{IEEEeqnarray*}
where $\epsilon^{(k)}_{j},\epsilon^{(k)}_{l,b},~\forall j \in \{1,2\},~\forall l,~\forall b,$ are positive tolerances, and $N_{\mathcal{C}}(\mathcal{X})$ and $N_{\overline{\mathcal{C}}}(\overline{\mathcal{X}})$ are the general normal cone of $\mathcal{C}$ and $\overline{\mathcal{C}}$, at points $\mathcal{X}$ and $\overline{\mathcal{X}}$, respectively.          
On the other hand, the optimality conditions of the ADMM blocks 1 and 2, problems \eqref{eq:Block 1} and \eqref{eq:Block 2}, result in:
\begin{IEEEeqnarray}{lCl}\label{Block 1 and 2 Optimality}
\small
0\in \nabla f(\mathcal{X}^{(t)}) +\sum_{b\in \mathcal{B}}\sum_{l\in\mathcal{L}} \boldsymbol{\psi}^{(t-1)}_{l,b} + \sum_{b\in \mathcal{B}} \Bigl[\rho^{(k)}_{s,b} \bigl(\mathbf{s}^{(t)}_b - \overline{\mathbf{s}}^{(t-1)}_b + \mathbf{z}^{(t-1)}_{s,b}\bigr) + \rho^{(k)}_{I,b} \bigl(\mathbf{I}^{(t)}_b - \overline{\mathbf{I}}^{(t-1)}_b + \mathbf{z}^{(t-1)}_{I,b}\bigr)\Bigr] 
+ N_{\mathcal{C}}(\mathcal{X}^{(t)}),~\qquad
\IEEEyesnumber \IEEEyessubnumber*\\[1ex]
\nonumber\\ 
0 \in -\sum_{b\in \mathcal{B}}\sum_{l\in\mathcal{L}} \boldsymbol{\psi}^{(t-1)}_{l,b} - \sum_{b\in \mathcal{B}} \Bigl[\rho^{(k)}_{s,b} \bigl(\mathbf{s}^{(t)}_b - \overline{\mathbf{s}}^{(t)}_b + \mathbf{z}^{(t-1)}_{s,b}\bigr) + \rho^{(k)}_{I,b} \bigl(\mathbf{I}^{(t)}_b - \overline{\mathbf{I}}^{(t)}_b + \mathbf{z}^{(t-1)}_{I,b}\bigr)\Bigr]
+ N_{\overline{\mathcal{C}}}(\overline{\mathcal{X}}^{(t)}).
\end{IEEEeqnarray}
% \begin{IEEEeqnarray}{lCl}\label{Block 1 and 2 Optimality}
% \small
% &0&~\in \nabla f(\mathcal{X}^{(t)}) +\sum_{b\in \mathcal{B}}\sum_{l\in\mathcal{L}} \boldsymbol{\psi}^{(t-1)}_{l,b}
%  + \sum_{b\in \mathcal{B}} \Bigl[\rho^{(k)}_{s,b} \bigl(\mathbf{s}^{(t)}_b - \overline{\mathbf{s}}^{(t-1)}_b + \nonumber \\
%  &&\mathbf{z}^{(t-1)}_{s,b}\bigr) 
% + \rho^{(k)}_{I,b} \bigl(\mathbf{I}^{(t)}_b - \overline{\mathbf{I}}^{(t-1)}_b + \mathbf{z}^{(t-1)}_{I,b}\bigr)\Bigr] 
% + N_{\mathcal{C}}(\mathcal{X}^{(t)}), \qquad 
% \IEEEyesnumber \IEEEyessubnumber*
% \end{IEEEeqnarray}
% \begin{IEEEeqnarray}{lCl}
% \small
% 0 &\in& -\sum_{b\in \mathcal{B}}\sum_{l\in\mathcal{L}} \boldsymbol{\psi}^{(t-1)}_{l,b}
%  - \sum_{b\in \mathcal{B}} \Bigl[\rho^{(k)}_{s,b} \bigl(\mathbf{s}^{(t)}_b - \overline{\mathbf{s}}^{(t)}_b + \mathbf{z}^{(t-1)}_{s,b}\bigr) 
% \nonumber\\
% && \quad\quad + \rho^{(k)}_{I,b} \bigl(\mathbf{I}^{(t)}_b - \overline{\mathbf{I}}^{(t)}_b + \mathbf{z}^{(t-1)}_{I,b}\bigr)\Bigr]
% + N_{\overline{\mathcal{C}}}(\overline{\mathcal{X}}^{(t)}).\IEEEyessubnumber*
% \end{IEEEeqnarray}

Considering the Lagrange variable update of the inner algorithm, the optimality conditions of ADMM blocks 1 and 2 can be expressed as
\begin{IEEEeqnarray}{lcl}\label{Optimality_Conditions}
\small
\sum_{b\in \mathcal{B}} \Bigl[
\rho^{(k)}_{s,b} \bigl( -\overline{\mathbf{s}}^{(t)}_b + \mathbf{z}^{(t)}_{s,b} 
+ \overline{\mathbf{s}}^{(t-1)}_{b} - \mathbf{z}^{(t-1)}_{s,b} \bigr)  + \rho^{(k)}_{I,b} \bigl( -\overline{\mathbf{I}}^{(t)}_b + \mathbf{z}^{(t)}_{I,b} + \overline{\mathbf{I}}^{(t-1)}_{b} - \mathbf{z}^{(t-1)}_{I,b} \bigr)
\Bigr]  \in \nabla f(\mathcal{X}^{(t)})  
+ \sum_{b\in \mathcal{B}} \sum_{l\in\mathcal{L}} \boldsymbol{\psi}^{(t)}_{l,b}
+ N_{\mathcal{C}}(\mathcal{X}^{(t)}),\nonumber\\*
\IEEEyesnumber\IEEEyessubnumber \\ 
\sum_{b\in \mathcal{B}} \sum_{l\in \mathcal{L}}\Bigl[
\rho^{(k)}_{l,b} \bigl(\mathbf{z}^{(t-1)}_{l,b} - \mathbf{z}^{(t)}_{l,b}\bigr)\Bigr] \in -\sum_{b\in \mathcal{B}} \sum_{l\in\mathcal{L}} \boldsymbol{\psi}^{(t)}_{l,b} + N_{\overline{\mathcal{C}}}(\overline{\mathcal{X}}^{(t)}).\nonumber\\*
\IEEEyessubnumber
\end{IEEEeqnarray}

Comparing \eqref{Optimality_Conditions} and \eqref{Stationary Point Conditions}, the stopping criteria for inner-level algorithm \ref{alg:Inner ADMM_Proposed} will be according to \eqref{eq:Inner_Stopping_Criteria}.
\subsubsection{Stopping criteria for outer-level Algorithm \ref{alg:alg_Distributed_Proposed}}
The outer-level Algorithm~\ref{alg:alg_Distributed_Proposed} aims at implementing ALM to move $\mathbf{z}_{l,b},~\forall b,~\forall l$, to zero. Furthermore, upon convergence, each BS will consider the beamforming weight matrices $\mathbf{W}^b,~\forall b,$ as the final decision. As a result, we consider the outer-level ALM algorithm to stop when the $\mathbf{z}_{l,b}$ values are small enough and the increase in the PF objective function value is less than a threshold, i.e.,:
\begin{IEEEeqnarray}{lCl}
    \left\| \mathbf{z}^{(k)}_{l,b}\right\|\leq \epsilon_{o,1},~\forall l,~\forall b, \qquad \IEEEyesnumber \IEEEyessubnumber*\\
    \frac{\big| f(\mathcal{X}^{(k)})- f(\mathcal{X}^{(k-1)}) \big|}{\big|f(\mathcal{X}^{(k-1)}) \big|} < \epsilon_{{o,2}}.
\end{IEEEeqnarray}
\subsection{Computational Complexity and Signaling Overhead}
In this subsection, we compare the per-iteration computational complexity of the centralized and distributed algorithms. For the distributed algorithm, we focus on the local optimization cost of the ADMM block 2, which is the dominant contributor to the per-iteration complexity. Specifically, ADMM block 2 subproblems take the form of SOC programs (SOCP) with constraints such as power limits and interference bounds that are SOC representable. The worst-case complexity of a primal-dual interior-point method for SOCP, as implemented in solvers such as \texttt{mosek} or \texttt{SeDuMi}~\cite{CVX,gb08,BoydBook}, scales approximately as $\mathcal{O}(n^{3.5})$, where $n$ denotes the total cone dimension. For ADMM Block~2, each BS $b$ solves a local SOCP whose total cone dimension is approximately $2U^2 + 2N_b U$. Therefore, the per-iteration worst-case computational complexity at BS~$b$ is approximately $\mathcal{O}\bigl((U^2 + N_b U)^{3.5}\bigr).$
Considering HAPS with $N_{B+1}$ antenna elements, typically much larger than that of an MBS, the per-iteration complexity of the distributed approach is therefore upper bounded by $\mathcal{O}\bigl((U^2 + N_{B+1} U)^{3.5}\bigr)$. In contrast, the centralized algorithm jointly optimizes over all BSs. Its problem dimension aggregates all local variables, leading to a total cone dimension proportional to $\sum_{b\in \mathcal{B}} (U^2+ N_b U)$. The resulting worst-case complexity per iteration is approximately $\mathcal{O}\Bigl(\bigl(\sum_{b\in \mathcal{B}} (U^2+ N_b U)\bigr)^{3.5}\Bigr),$ which scales much worse in the network size, especially as the number of BSs or antenna elements grows. Furthermore, the distributed approach offers significant improvements in signaling overhead. In the centralized scheme, each BS must transmit all local CSI to a central controller, amounting to $2 N_b U$ real scalars per BS. By contrast, in the distributed scheme, only the consensus variables need to be exchanged, requiring approximately $3U$ real scalars per BS. This reduced overhead is particularly advantageous when $N_b$ is large, specifically in the HAPS case.
\section{Numerical Results and Discussion}\label{Sec:Results}
\begin{table}[!t]
\caption{Simulation Parameters.}\label{tab:SimulationParameters}
\centering
\begin{tabular}{|c||c|}
\hline
\textbf{Parameter} & \textbf{Value}\\
\hline
HAPS altitude & $20$ Km~
\cite{itu_wrc23}\\
\hline
Center frequency ($f_c$) & $2.545$ GHz~\cite{itu_wrc23}\\
\hline
$\sigma_\xi$,~$K_u$  & $8$,~$10$~\cite{HAPS-MIMO}\\
\hline
$d_x,~d_y$ & $\lambda/2$\\
\hline
$\sigma^2_n$ & $-100$ dBm\\
\hline
$N_{\text{b}},~\forall b \in \mathcal{B}\setminus\{B+1\},~N_{B+1}$ & $4 \times 4,~8 \times 8$\\
\hline
$P_b^{\text{max}},~\forall b\in \mathcal{B}\setminus\{B+1\},~P_{B+1}^{\text{max}}$ & $43$ dBm,~$52$ dBm~\cite{itu_wrc23}\\
\hline
Outer-level penalty parameter ($\rho^{(0)}_o$) & $10$\\
\hline
$\delta,~\omega,~\gamma$ & $2,~0.5,~1.5$~\cite{TWoStepDistributed}\\
\hline
\end{tabular}
\end{table}
\begin{figure}[t]
    \centering
    \begin{minipage}{\linewidth}
        \centering
        \includegraphics[width=0.6\linewidth]{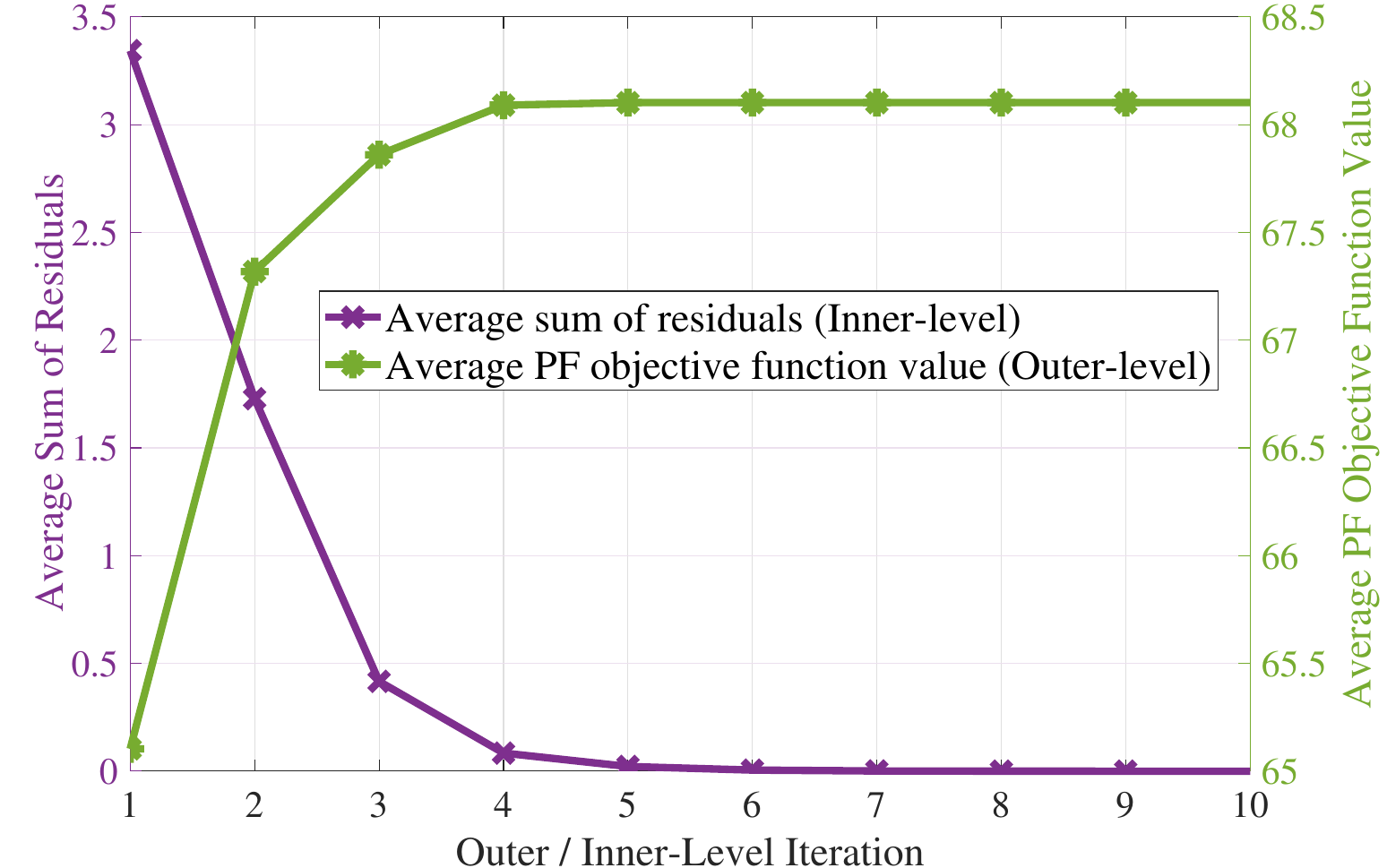}
        \captionof{figure}{\small Convergence behavior of the proposed PFBWD algorithm.}
        \label{fig:Convergence}
    \end{minipage}
\end{figure}

\begin{table}[!t]
\caption{\small Impact of inner-level penalty parameter on the required number of iterations.}\label{tab:deltaImpact}
\centering
\begin{tabular}{|c||c||c||c|}
\hline
\textbf{Parameter} & \textbf{$\delta$=0.5}& \textbf{$\delta$=1}& \textbf{$\delta$=2}\\
\hline
Inner-level algorithm iterations & $13.5$ & $12.5$ & $8.33$\\
\hline
\end{tabular}
\end{table}
In this section, we evaluate the performance of the proposed distributed PFBWD algorithm in HAPS-empowered vHetNets and compare it against both standalone terrestrial networks and the centralized approach. The standalone terrestrial network refers to a scenario where no HAPS is involved. Integrating HAPS into terrestrial networks under a harmonized spectrum introduces strong inter-tier interference, which raises concerns regarding potential performance degradation compared to standalone terrestrial networks. Therefore, it is essential to benchmark the proposed system model against a standalone terrestrial network. \textcolor{black}{Moreover, comparisons with the centralized algorithm are necessary since it serves as an upper performance bound, enabling a clear assessment of the optimality loss incurred by distributed implementation.} In the case of a centralized algorithm, the original problem is solved using the SCA framework, which incorporates some RLT, as detailed in~\cite{Our-WCL}. The simulated network consists of four MBSs and one HAPS, deployed over a $4~\text{km} \times 4~\text{km}$ square urban area. The HAPS and MBSs are equipped with $8 \times 8$ and $4 \times 4$ MIMO antenna arrays, respectively, and jointly serve $U$ single-antenna UEs in a cell-free architecture. The UEs are uniformly distributed across the considered area. The remaining simulation parameters and their values are summarized in Table~\ref{tab:SimulationParameters}. All results are averaged over $1000$ independent network realizations. The algorithm is implemented using CVX parser, which runs on MATLAB, with \texttt{mosek} 9.1.9 as an internal solver~\cite{CVX}.

First, we discuss the convergence behavior of the proposed distributed PFBWD algorithm. Fig.~\ref{fig:Convergence} plots the average sum of residuals, defined as $\sum_{b\in\mathcal{B}}\sum_{l\in\mathcal{L}}\left\|\mathbf{r}_{l,b}\right\|$, over the inner-level ADMM Algorithm~\ref{alg:Inner ADMM_Proposed} iterations $t$, and the average PF objective function value (i.e., $\sum_{u\in\mathcal{U}}{\log(\log_2(1+\gamma_u))}$) over the outer-level Algorithm~\ref{alg:alg_Distributed_Proposed} iterations $k$. It can be observed that the residual values converge to zero as $t$ increases, while the objective value increases with $k$, demonstrating the convergence behavior of the proposed algorithm. The outer-level ALM algorithm also exhibits fast convergence as the objective value saturates after a few iterations (on average $k=4$ iterations). However, the choice of penalty parameters plays a critical role in determining the number of required iterations. As shown in Table~\ref{tab:deltaImpact}, as the ratio of the inner-level to outer-level penalty parameters (i.e., $\delta$ in Algorithm~\ref{alg:Inner ADMM_Proposed}) increases, the number of required iterations reduces. This is because a higher penalty parameter enforces the satisfaction of the coupling constraints more aggressively, thereby accelerating convergence to feasibility. However, excessively large values of $\delta$ may lead to numerical instability~\cite{BoydADMM}.

\begin{figure*}[t]
    \centering
    \begin{subfigure}[b]{0.47\linewidth}
        \centering
        \includegraphics[width=\linewidth]{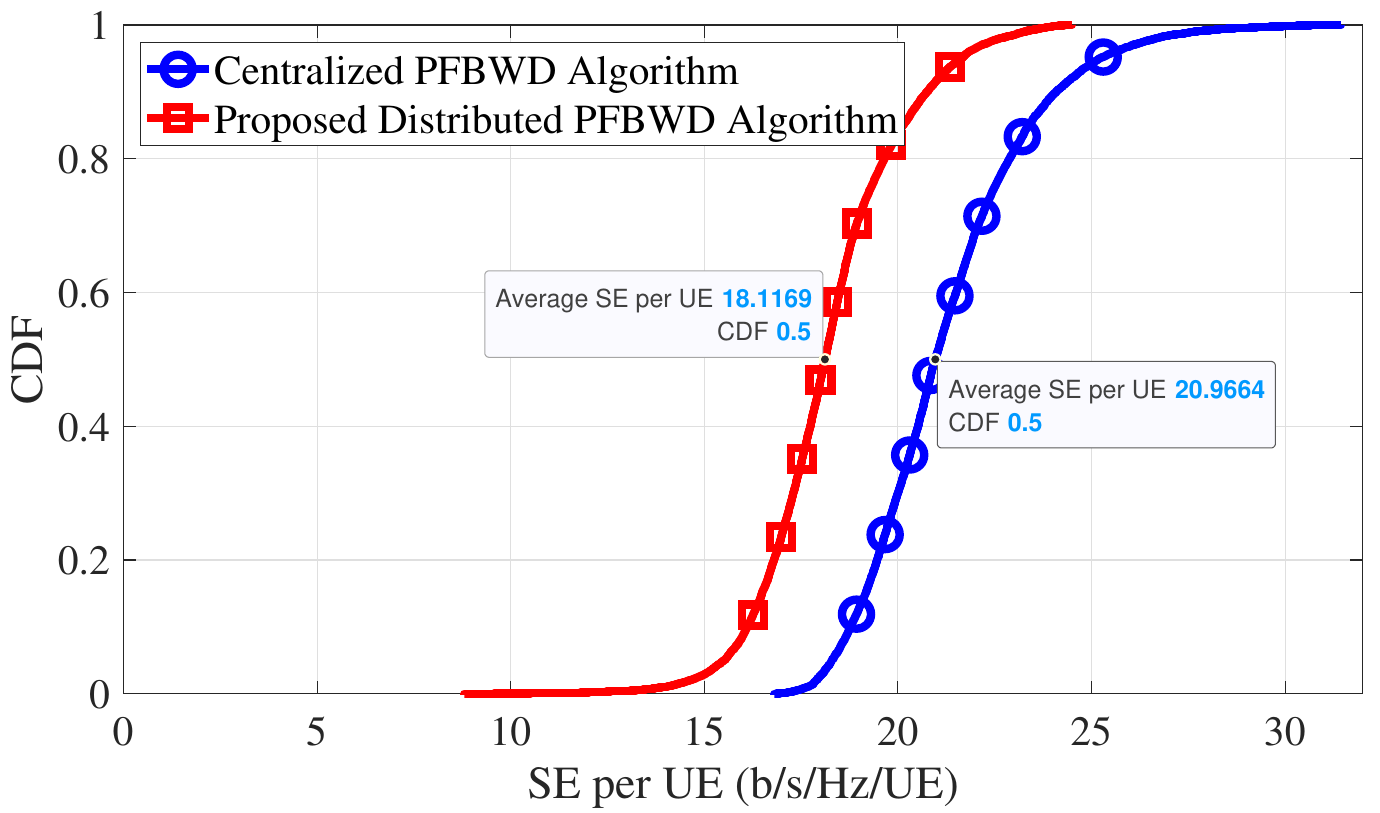}
        \caption{\small CDF of SE per UE (b/s/Hz/UE).}
        \label{fig:se_cdf}
    \end{subfigure}
    \hfill
    \begin{subfigure}[b]{0.47\linewidth}
        \centering
        \includegraphics[width=\columnwidth]{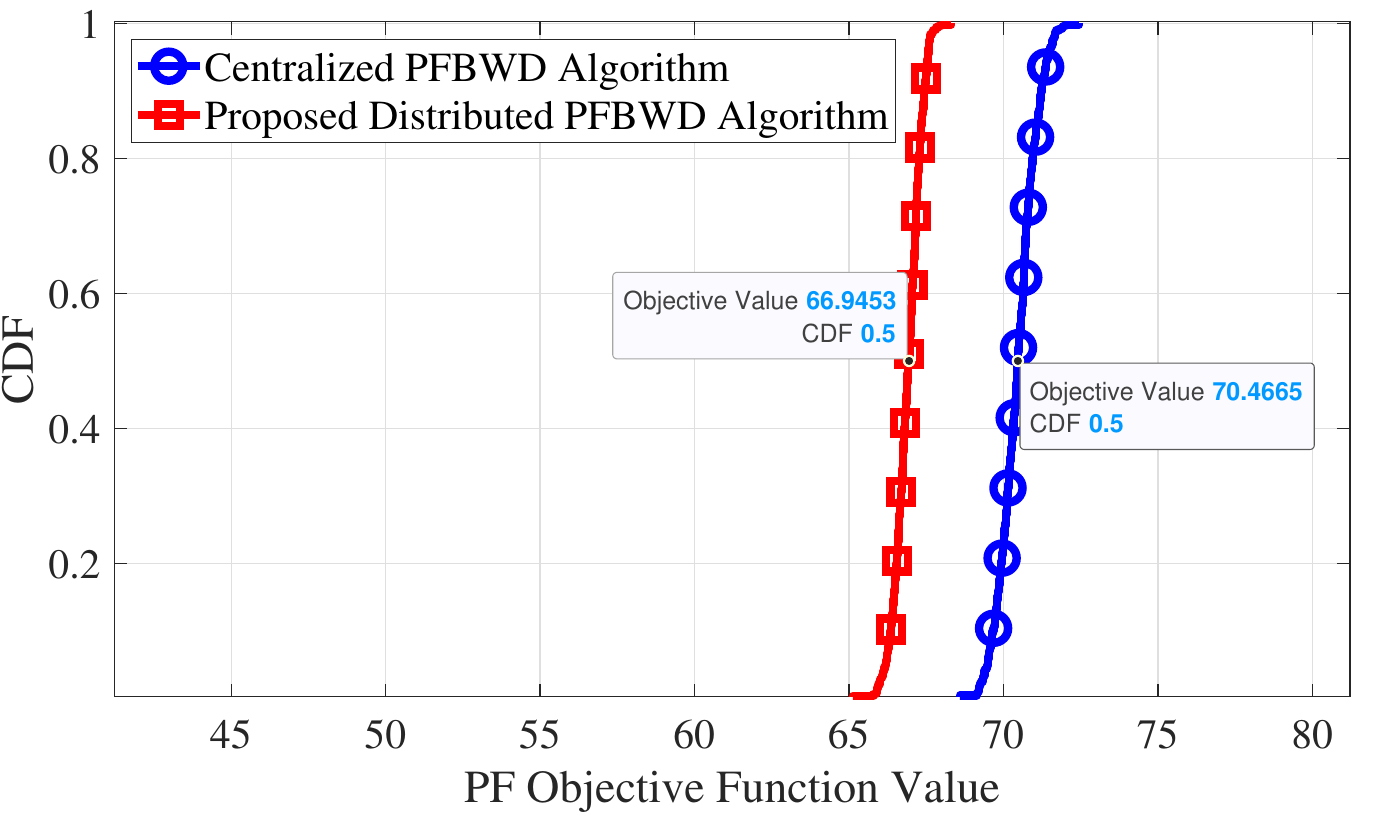}
        \caption{\small CDF of PF objective function value.}
        \label{fig:obj_cdf}
    \end{subfigure}
    \caption{\small Statistical behavior of SE per UE and PF objective function value in a HAPS-empowered vHetNet ($4$ MBSs + $1$ HAPS) serving $16$ UEs.}
    \label{fig:statistical_behavior}
\end{figure*}
\begin{table}[!t]
\centering
\caption{\small Comparison of computational complexity and signaling overhead for centralized vs. distributed PFBWD algorithms in a vHetNet ($4$ MBSs + $1$ HAPS) serving $16$ UEs.}
\label{tab:table2}
\centering
\begin{tabular}{|c||c|c|}
\hline
\textbf{Parameter} & \textbf{Distributed} & \textbf{Centralized} \\
\hline
Computational complexity
& $\mathcal{O}(1280^{3.5})$
& $\mathcal{O}(3328^{3.5})$ \\
\hline
\makecell{Signaling overhead \\ (real scalars per BS)} 
& $\approx 48$
& \makecell{$\approx 512$ for MBS\\ $\approx 2048$ for HAPS} \\
\hline
\end{tabular}
\end{table}

\begin{figure}[t]
    \centering
    \begin{minipage}{\linewidth}
        \centering
        \includegraphics[width=0.6\linewidth]{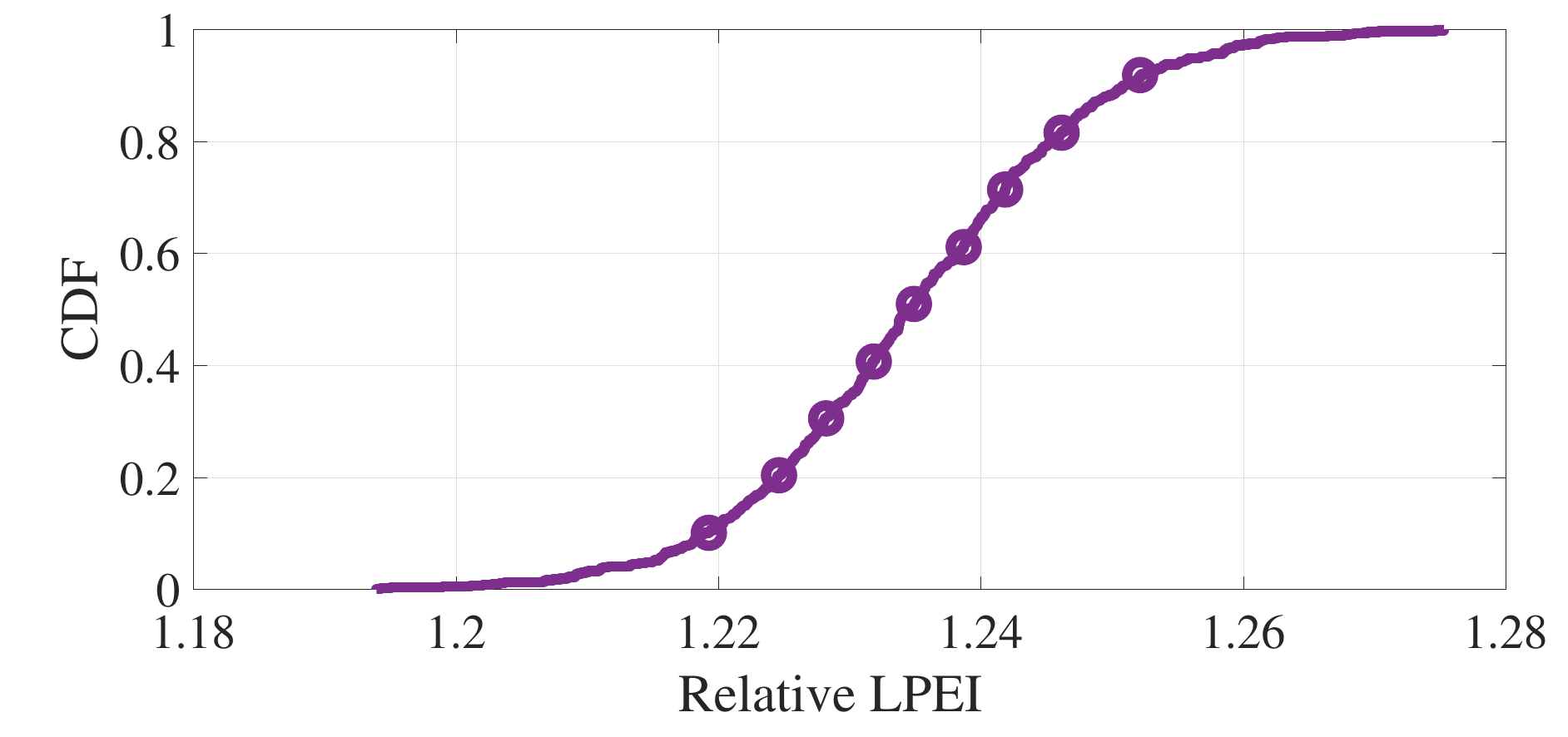}
        \captionof{figure}{\small CDF of relative LPEI~(proposed distributed PFBWD algorithm~\ref{alg:alg_Distributed_Proposed} over centralized algorithm).}
        \label{fig:LPEI}
    \end{minipage}
\end{figure}

Table~\ref{tab:table2} summarizes the computational complexity and signaling overhead requirements of the centralized and distributed algorithms when serving $16$ UEs, considering $4 \times 4$ and $8 \times 8$ antenna arrays at the MBSs and the HAPS, respectively. It is evident that the proposed distributed PFBWD Algorithm~\ref{alg:alg_Distributed_Proposed} incurs substantially lower computational complexity and significantly reduced data exchange, resulting in a pronounced reduction in signaling overhead compared to the centralized benchmark.
Furthermore, Figs.~\ref{fig:se_cdf} and~\ref{fig:obj_cdf} depict the cumulative distribution functions (CDFs) of SE per UE and the PF objective value, respectively, for a vHetNet comprising $4$ MBSs and $1$ HAPS serving $16$ UEs. As expected, the centralized algorithm achieves higher performance due to its access to global CSI and centralized optimization. In contrast, the distributed algorithm relies solely on local CSI and limited coordination among BSs.
Quantitatively, Fig.~\ref{fig:se_cdf} shows that the average SE per UE achieved by the centralized algorithm is approximately $20.9$~b/s/Hz/UE, whereas the distributed algorithm attains $18.1$~b/s/Hz/UE, corresponding to a performance gap of about $13\%$. Similarly, Fig.~\ref{fig:obj_cdf} indicates average PF objective values of $70.5$ and $66.9$ for the centralized and distributed schemes, respectively, resulting in a gap of approximately $5\%$. The similar slopes and spreads of the CDFs indicate comparable variability across network realizations, confirming that the distributed algorithm operates close to the centralized benchmark in terms of achievable performance.

To jointly evaluate the achieved performance in terms of the objective function value, computational cost, and the signaling overhead, we define a logarithmic performance--efficiency index (LPEI) as
\begin{equation}
\small
\text{LPEI}
=
\frac{\text{Average Objective Function Value}}{\log_{10}(\text{Computational Complexity} \times \text{Signaling Overhead})}.
\end{equation}

Logarithmic normalization is applied to ensure meaningful comparison when costs between the distributed and centralized algorithms differ significantly.
Accordingly, Fig.~\ref{fig:LPEI} plots the CDF of the relative LPEI, defined as the ratio between the LPEI of the proposed distributed PFBWD algorithm and that of the centralized one. As observed, the relative LPEI remains consistently greater than one across all network realizations, indicating that the proposed distributed PFBWD algorithm is uniformly more efficient than the centralized solution. These results confirm that, despite a modest reduction in SE and PF performance, the proposed distributed PFBWD algorithm achieves higher overall efficiency by substantially reducing computational complexity and signaling overhead, making it a scalable and practical solution for large-scale HAPS-empowered vHetNets.

\begin{figure*}[t]
    \centering
    \begin{subfigure}[t]{0.48\textwidth}
        \centering
        \includegraphics[width=\linewidth]{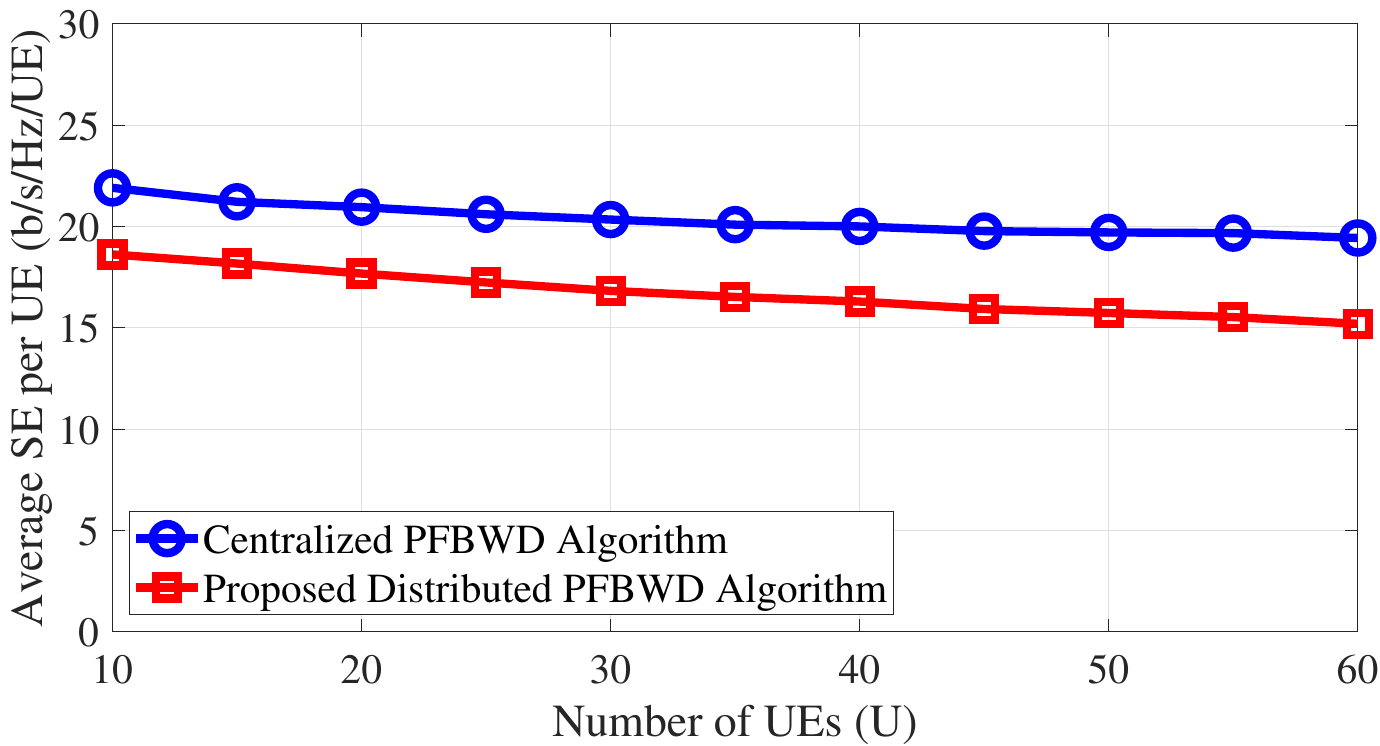}
        \caption{\small Average SE per UE vs. number of UEs}
        \label{fig:SEvsNumUE:a}
    \end{subfigure}
    \hfill
    \begin{subfigure}[t]{0.48\textwidth}
        \centering
        \includegraphics[width=\linewidth]{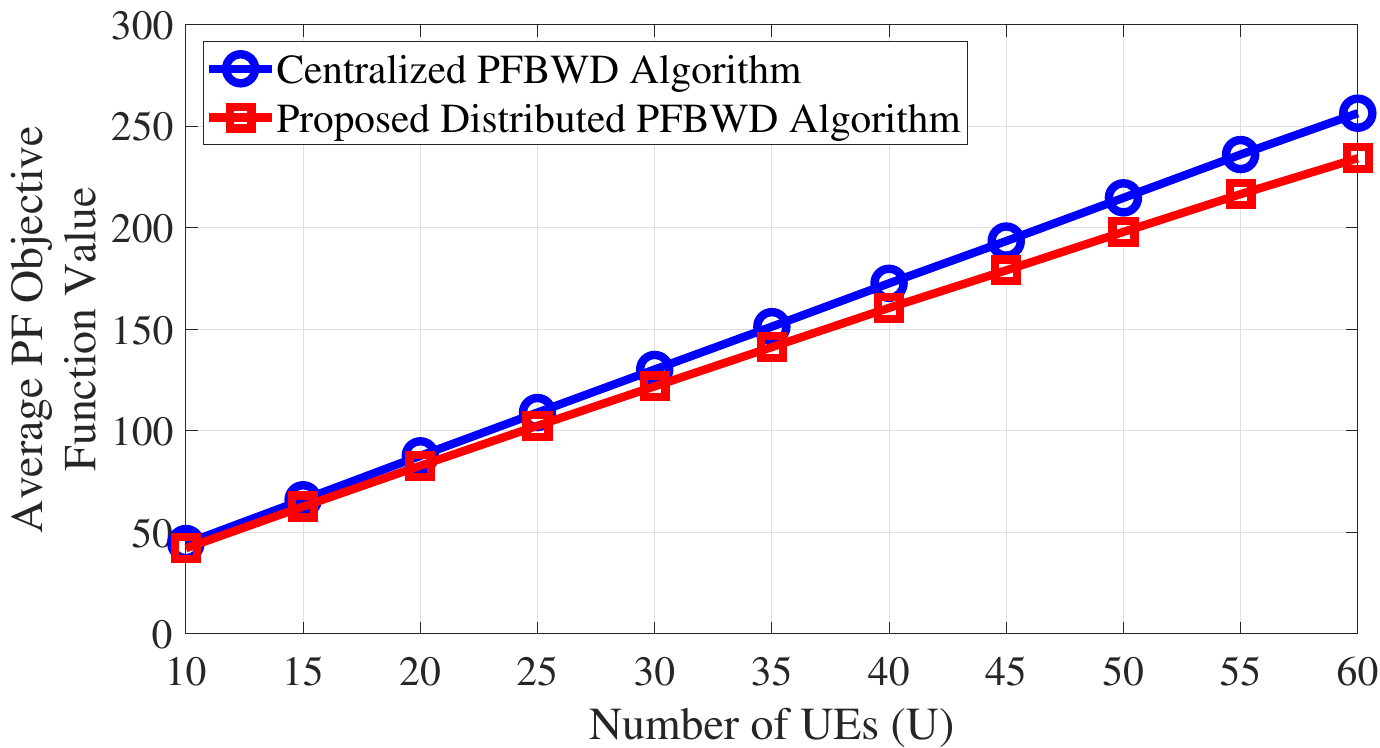}
        \caption{\small Average PF objective function value vs. number of UEs}
        \label{fig:SEvsNumUE:b}
    \end{subfigure}
    \caption{\small Performance comparison of centralized and distributed PFBWD algorithms for different numbers of UEs in a HAPS-empowered vHetNet ($4$ MBSs + $1$ HAPS).}
    \label{fig:SEvsNumUE}
\end{figure*}

\color{black}

Furthermore, Fig.~\ref{fig:SEvsNumUE} illustrates the performance gap between the centralized and distributed PFBWD algorithms with an increasing number of UEs. Specifically, Figs.~\ref{fig:SEvsNumUE:a} and ~\ref{fig:SEvsNumUE:b} plot the average SE per UE and the average PF objective function value, respectively, in a HAPS-empowered vHetNet consisting of $4$ MBSs and $1$ HAPS. As the number of UEs grows, the average SE per UE decreases mainly due to two reasons: i) the available resources are shared among more UEs and ii) the increased interference. Additionally, the performance gap between the centralized and distributed algorithms slightly widens with increasing UEs. This is because the amount of missing information at each BS in the distributed approach grows, leading to a minor decline in performance. Importantly, this minor decrease in performance comes with significant gains in algorithmic complexity and reduced signaling overhead among BSs. 
\begin{figure}[t]
    \centering
    \begin{minipage}{\linewidth}
        \centering
        \includegraphics[width=0.6\linewidth]{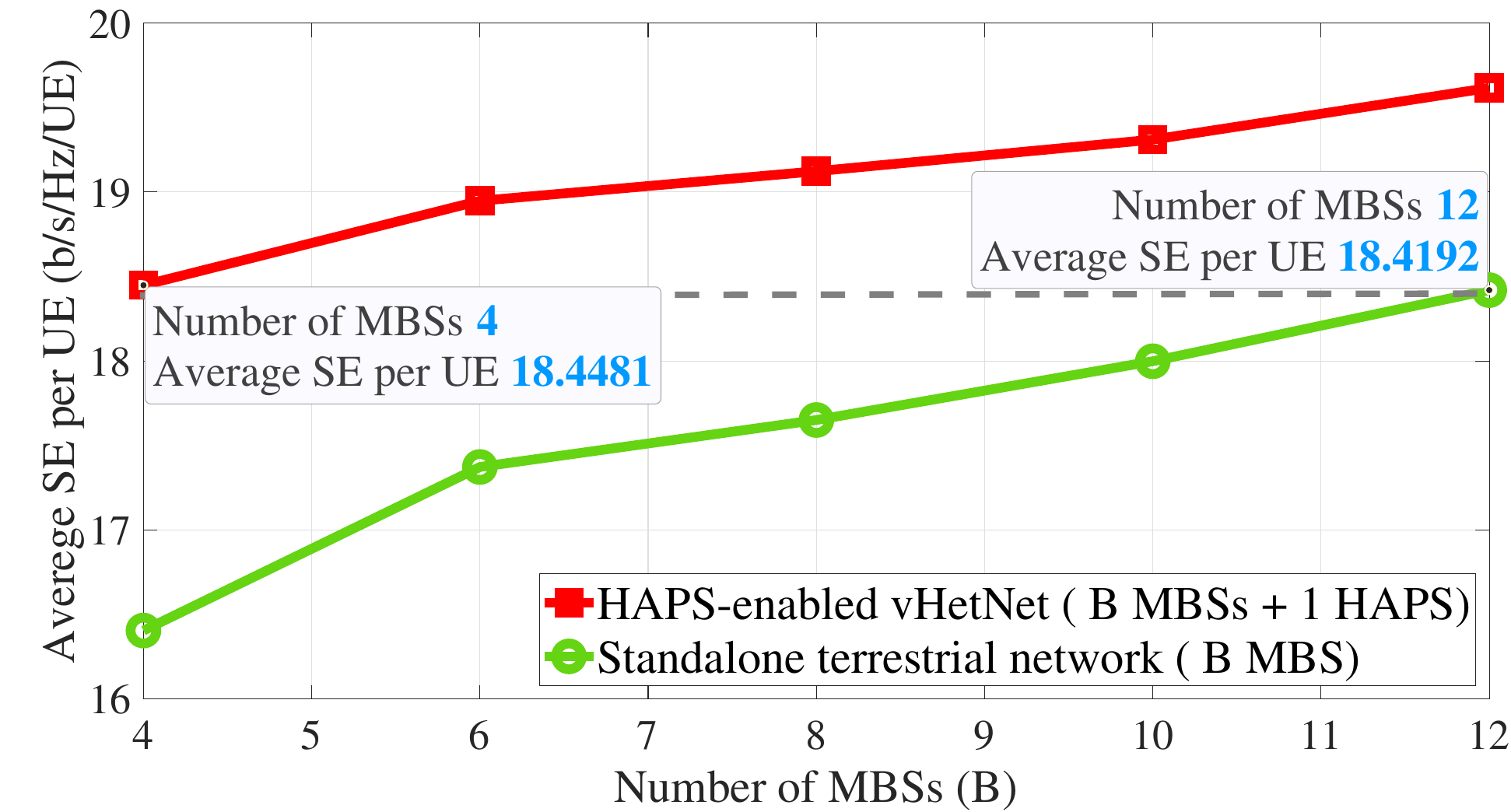}
        \captionof{figure}{\small Average SE per UE: vHetNet vs. standalone terrestrial network across different numbers of MBSs.}
        \label{fig:vHetNetvsTerrestrial}
    \end{minipage}
\end{figure}

To further examine the impact of vHetNet architecture and HAPS capabilities on network performance, Fig.~\ref{fig:vHetNetvsTerrestrial} compares the performance of the HAPS-empowered vHetNet and the standalone terrestrial network serving $16$ UEs, under the proposed PFBWD algorithm. First, it can be observed that increasing the number of MBSs in both vHetNet and standalone terrestrial networks improves the average SE per UE, primarily due to the availability of more resources, including a higher number of antenna elements. Second, integrating HAPS into the terrestrial network architecture yields a noticeable enhancement in the average SE per UE. Specifically, for all values of $B$, integrating the terrestrial network with one HAPS, resulting in a total of $(B+1)$ BSs, consistently leads to improved SE for UEs. Furthermore, it can be observed that to match the performance of the $(4+1)$ vHetNet configuration ($4$ MBSs and $1$ HAPS), the standalone terrestrial network requires a total of $12$ MBSs. In other words, the vHetNet with $4$ MBSs and $1$ HAPS achieves the same average SE per UE as a standalone terrestrial network comprising $12$ MBSs, clearly demonstrating the efficiency of integrating HAPS. Importantly, these results also highlight the scalability of the proposed distributed algorithm, which can efficiently solve the beamforming design problem even for large numbers of MBSs in the network.

\begin{figure}[t]
    \centering
    \begin{minipage}{\linewidth}
        \centering
        \includegraphics[width=0.6\linewidth]{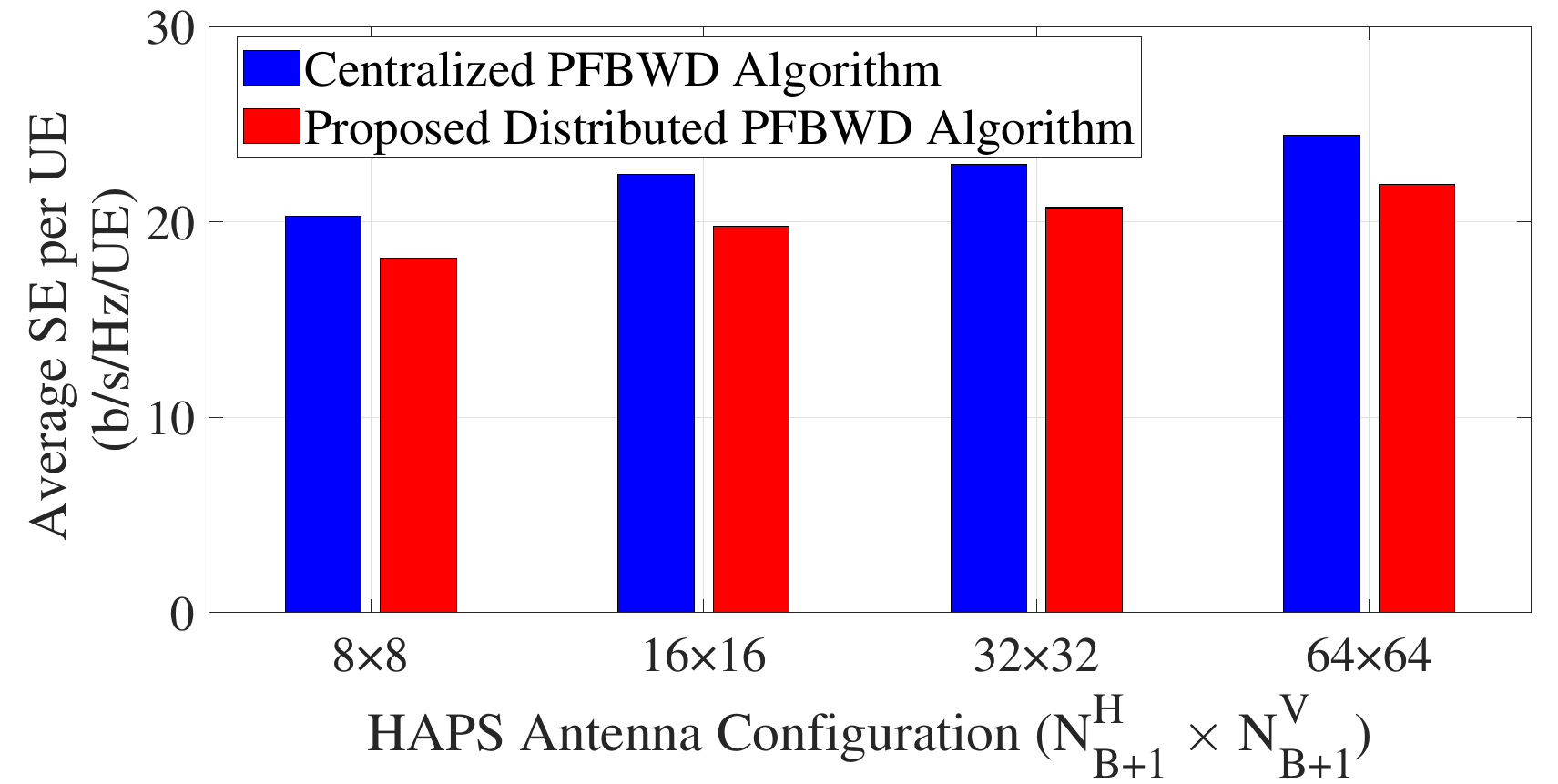}
        \captionof{figure}{\small Impact of HAPS antenna architecture on the average SE per UE in a vHetNet ($4$ MBSs + $1$ HAPS) serving $16$ UEs.}
        \label{fig:HAPSAntennaResult}
    \end{minipage}
\end{figure}

\begin{figure}[t]
    \centering
    \begin{minipage}{\linewidth}
        \centering
        \includegraphics[width=0.6\linewidth]{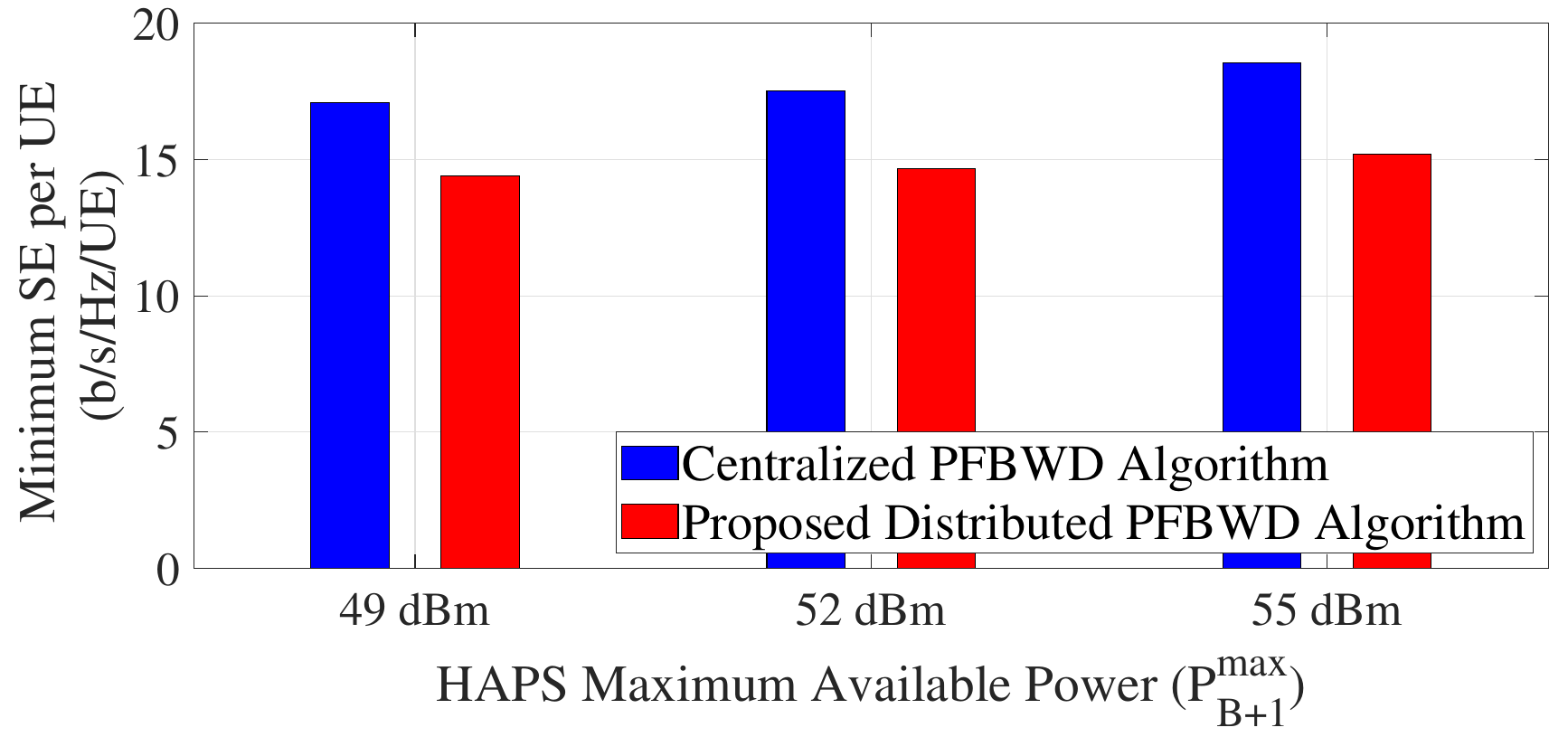}
        \captionof{figure}{\small Impact of ($P^{\text{max}}_{B+1}$) on the minimum SE per UE in a vHetNet ($4$ MBSs + $1$ HAPS) serving $16$ UEs.}
        \label{fig:HAPSPower}
    \end{minipage}
\end{figure}

In addition to the network architecture and optimization framework, the system-level parameters of the HAPS play a crucial role in determining the achievable SE in HAPS-empowered vHetNets. In particular, the antenna configuration and maximum available transmit power at the HAPS directly influence the beamforming capability, interference management effectiveness, and overall performance. To this end, Fig.~\ref{fig:HAPSAntennaResult} presents the average SE per UE in a $(4+1)$ vHetNet configuration serving $16$ UEs, for various HAPS antenna configurations. As shown, the SE gain provided by HAPS integration can be further amplified by increasing the number of antenna elements at the HAPS. This is particularly promising, as the large surface area of HAPS platforms allows for accommodating a significantly larger number of antennas, enabling improved beamforming capabilities and higher spatial resolution.
Furthermore, Fig.~\ref{fig:HAPSPower} illustrates the minimum SE per UE for different maximum available transmit power, i.e., $P^{\text{max}}_{B+1}$, levels at the HAPS. The results indicate that increasing the HAPS transmit power leads to improvement in the achieved minimum SE, highlighting the importance of power provisioning at the HAPS for enhancing fairness and ensuring satisfactory service to UEs.

\textcolor{black}{Building on the promising performance demonstrated in this study, several open research directions in cell-free HAPS-empowered vHetNets remain to be explored. These include the impact of imperfect CSI, particularly at the HAPS tier, synchronization challenges between the HAPS and terrestrial network tiers, and the limited wireless backhaul capacity of the HAPS tier, especially when determining the number of associated UEs and developing scalable resource allocation strategies.}
\section{Conclusion}\label{Sec:Conclusion}
This paper addressed the challenges of interference management in harmonized spectrum HAPS-empowered vertical heterogeneous networks (vHetNets). We highlighted the limitations of centralized interference management approaches, which face scalability issues due to the large number of MBSs, high-dimensional antenna arrays, and significant signaling overhead. To overcome these challenges, we formulated a cell-free proportional fairness beamforming weight design (PFBWD) problem tailored for HAPS-empowered vHetNet deployments. Recognizing the nonconvex nature of the problem, we adopted a two-level distributed algorithm that embeds a structured three-block ADMM within an augmented Lagrangian framework, ensuring guaranteed convergence. Simulation studies confirm convergence, effectiveness and scalability of the algorithm, showing that the algorithm can efficiently handle large-scale network configurations with many MBSs and high-dimensional antenna arrays. These findings underscore the practical value of advanced distributed optimization methods for enabling efficient and scalable interference management in future integrated aerial-terrestrial networks.
\bibliographystyle{IEEEtran}
\bibliography{Ref}

@article{BoydADMM,
author = {Boyd, Stephen and Parikh, Neal and Chu, Eric and Peleato, Borja and Eckstein, Jonathan},
title = {Distributed Optimization and Statistical Learning via the Alternating Direction Method of Multipliers},
year = {2011},
publisher = {Now Publishers Inc.},
address = {Hanover, MA, USA},
volume = {3},
number = {1},
issn = {1935-8237},
url = {https://doi.org/10.1561/2200000016},
doi = {10.1561/2200000016},
journal = {Found. Trends Mach. Learn.},
month = {Jan.},
pages = {1–122},
numpages = {122}
}

@ARTICLE{HAPSSurvey,
  author={Karabulut Kurt, Gunes and Khoshkholgh, Mohammad G. and Alfattani, Safwan and Ibrahim, Ahmed and Darwish, Tasneem S. J. and Alam, Md Sahabul and Yanikomeroglu, Halim and Yongacoglu, Abbas},
  journal={IEEE Commun. Surveys Tuts.}, 
  title={A Vision and Framework for the High Altitude Platform Station ({HAPS}) Networks of the Future}, 
  year={2021},
  volume={23},
  number={2},
  pages={729-779},
month={Secondquarter}}

@ARTICLE{HAPS-MIMO,
  author={Lian, Zhuxian and Jiang, Lingge and He, Chen and He, Di},
  journal={IEEE Wireless Commun. Lett.}, 
  title={User Grouping and Beamforming for {HAP} Massive {MIMO} Systems Based on Statistical-Eigenmode}, 
  year={2019},
  volume={8},
  number={3},
  month={Jun.},
  pages={961-964},
  doi={10.1109/LWC.2019.2902140}}

@ARTICLE{Animesh-IEEEAccess,
  author={Yadav, Animesh and Dobre, Octavia A. and Ansari, Nirwan},
  journal={IEEE Access}, 
  title={Energy and Traffic Aware Full-Duplex Communications for {5G} Systems}, 
  year={2017},
  volume={5},
  number={},
  pages={11278-11290},
  doi={10.1109/ACCESS.2017.2696822}}

@book{BoydBook,
  title={Convex optimization},
  author={Boyd, Stephen and Vandenberghe, Lieven},
  year={2004},
  publisher={Cambridge university press}
}

@misc{IMT-2030,
    title = {Framework and overall objectives of the future development of {IMT} for 2030 and beyond, {R}ecommendation {ITU-R M.}2160-0} ,
    author= {{International Telecommunications Union Radiocommunication Sector (ITU-R)}},
 address      = {Geneva, Switzerland},
    year = {2023}
}

@ARTICLE{Our-WCL,
  author={Alidadi Shamsabadi, Afsoon and Yadav, Animesh and Abbasi, Omid and Yanikomeroglu, Halim},
  journal={IEEE Wireless Commun. Lett.}, 
  title={Handling Interference in Integrated {HAPS}-Terrestrial Networks Through Radio Resource Management}, 
  year={2022},
  month={Dec.},
  volume={11},
  number={12},
  pages={2585-2589},
  doi={10.1109/LWC.2022.3210435}}

@ARTICLE{our-CL,
  author={Shamsabadi, Afsoon Alidadi and Yadav, Animesh and Yanikomeroglu, Halim},
  journal={IEEE Commun. Lett.}, 
  title={Enhancing Next-Generation Urban Connectivity: Is the Integrated {HAPS}-Terrestrial Network a Solution?}, 
  year={2024},
  month={May},
  volume={28},
  number={5},
  pages={1112-1116}}

@INPROCEEDINGS{our-ICC,
  author={Alidadi Shamsabadi, Afsoon and Yadav, Animesh and Yanikomeroglu, Halim},
  booktitle={2024 IEEE International Conference on Communications Workshops (ICC Workshops)}, 
  title={Impact of Objective Function on Spectral Efficiency in Integrated {HAPS}-Terrestrial Networks}, 
  year={2024},
  volume={},
  number={},
  pages={1895-1900},}

@article{TWoStepDistributed,
  author = {Kaizhao Sun and X. Andy Sun},
  title = {A two-level distributed algorithm for nonconvex constrained optimization},
  journal = {Computational Optimization and Applications},
  volume = {84},
  number = {2},
  pages = {609--649},
  year = {2023},
  month = {Mar.},
  doi = {10.1007/s10589-022-00433-4},
  url = {https://doi.org/10.1007/s10589-022-00433-4},
  issn = {1573-2894},
}

@ARTICLE{6GVTM,
  author={Shamsabadi, Afsoon Alidadi and Yadav, Animesh and Gadallah, Yasser and Yanikomeroglu, Halim},
  journal={IEEE Veh. Technol. Mag.}, 
  title={Exploring the {6G} Potentials: Immersive, Hyperreliable, and Low-Latency Communication}, 
  year={2025},
  volume={20},
  month={Mar.},
  number={1},
  pages={74-82},
  doi={10.1109/MVT.2025.3531346}}

@TechReport{itu_wrc,

	institution    ={International Telecommunication Union (ITU)},
	title         = {{World Radiocommunication Conference 2023 ({WRC}-23) Final Acts}},
	year          = {2023},
	url		= {https://www.itu.int/wrc-23/},
	month         = {December},
}

@ARTICLE{HAPSIM-TVT,
  author={Kawamoto, Yuichi and Okawara, Yuto and Verma, Shikhar and Kato, Nei and Kaneko, Kazuma and Sata, Ayaka and Ochiai, Mari},
  journal={IEEE Trans. Veh. Tech.}, 
  title={Interference Suppression in {HAPS}-based Space-Air-Ground Integrated Networks Using a Codebook-Based Approach}, 
  year={2024},
month={Dec.},
  volume={73},
  number={12},
  pages={19252-19262},
  doi={10.1109/TVT.2024.3446999}}

@ARTICLE{HAPS-Alouini,
  author={Belmekki, Baha Eddine Youcef and Aljohani, Abdulah Jeza and Althubaity, Saud A. and Harthi, Abdulhadi Al and Bean, Kevin and Aijaz, Adnan and Alouini, Mohamed-Slim},
  journal={IEEE Open J. Commun. Soc.}, 
  title={Cellular Network From the Sky: Toward People-Centered Smart Communities}, 
  year={2024},
  volume={5},
  number={},
  pages={1916-1936}}

@ARTICLE{HAPSIM-Cell,
  author={Shibata, Yohei and Takabatake, Wataru and Hoshino, Kenji and Nagate, Atsushi and Ohtsuki, Tomoaki},
  journal={IEEE Access}, 
  title={{HAPS} Cell Design Method for Coverage Extension Considering Coexistence on Terrestrial Mobile Networks}, 
  year={2024},
  volume={12},
  number={},
  pages={55506-55520},
  doi={10.1109/ACCESS.2024.3390116}}

@ARTICLE{SpectrumSharing,
  author={Ishikawa, Tsutomu and Tashiro, Koji and Konishi, Mitsukuni and Hoshino, Kenji},
  journal={IEICE Commun. Express}, 
  title={Spectrum Sharing in Integrated {HAPS} and Terrestrial Systems Using an Interference Canceler and Coordination}, 
  year={2024},
  volume={13},
  number={6},
  pages={185-189}}

@ARTICLE{multi-groupBF,
  author={Chen, Erkai and Tao, Meixia},
  journal={IEEE Trans. Commun.}, 
  title={{ADMM}-Based Fast Algorithm for Multi-Group Multicast Beamforming in Large-Scale Wireless Systems}, 
  year={2017},
month={Jun.},
  volume={65},
  number={6},
  pages={2685-2698},
  doi={10.1109/TCOMM.2017.2679708}}

@ARTICLE{StochasticADMM,
  author={Hu, Shuyan and Xu, Chongbin and Wang, Xin and Huang, Yongwei and Zhang, Shunqing},
  journal={IEEE Trans. Veh. Technol.}, 
  title={A Stochastic {ADMM} Approach to Distributed Coordinated Multicell Beamforming for Renewables Powered Wireless Cellular Networks}, 
  year={2018},
month={Sep.},
  volume={67},
  number={9},
  pages={8595-8607},
  doi={10.1109/TVT.2018.2853605}}

@article{jiang2019structured,
  title={Structured nonconvex and nonsmooth optimization: algorithms and iteration complexity analysis},
  author={Jiang, Bin and Lin, Ting Kei and Ma, Shiqian and Zhang, Shuzhong},
  journal={Computational Optimization and Applications},
  volume={72},
  number={1},
  pages={115--157},
  year={2019},
  publisher={Springer},
  doi={10.1007/s10589-018-0034-y}
}

@ARTICLE{FullDimension,
  author={Dong, Rui and Li, Ang and Hardjawana, Wibowo and Li, Yonghui and Ge, Xiaohu and Vucetic, Branka},
  journal={IEEE Trans. Veh. Technol.}, 
  title={Joint Beamforming and User Association Scheme for Full-Dimension Massive {MIMO} Networks}, 
  year={2019},
month={Aug.},
  volume={68},
  number={8},
  pages={7733-7746},
  doi={10.1109/TVT.2019.2923415}}

@ARTICLE{DistributedBFBP,
  author={Sohn, Illsoo and Lee, Sang Hyun and Andrews, Jeffrey G.},
  journal={IEEE Trans. Wireless Commun.}, 
  title={Belief Propagation for Distributed Downlink Beamforming in Cooperative {MIMO} Cellular Networks}, 
  year={2011},
month={Dec.},
  volume={10},
  number={12},
  pages={4140-4149},
  doi={10.1109/TWC.2011.101210.101698}}

@ARTICLE{Marzetta,
  author={Ngo, Hien Quoc and Ashikhmin, Alexei and Yang, Hong and Larsson, Erik G. and Marzetta, Thomas L.},
  journal={IEEE Trans. on Wireless Commun.}, 
  title={Cell-Free Massive {MIMO} Versus Small Cells}, 
  year={2017},
month={Mar.},
  volume={16},
  number={3},
  pages={1834-1850},}

@ARTICLE{IM_Magazine,
  author={Shamsabadi, Afsoon Alidadi and Yadav, Animesh and Yanikomeroglu, Halim},
  journal={IEEE Commun. Stand. Mag.}, 
  title={Interference Management Strategies for {HAPS}-Enabled {vHetNets} in Urban Deployments}, 
  year={2025},
month={Jun.},
  volume={9},
  number={2},
  pages={56-62},
  doi={10.1109/MCOMSTD.2025.3569011}}

@ARTICLE{MBSChannel,
  author={Ghafoor, Umar and Khan, Humayun Zubair and Ali, Mudassar and Siddiqui, Adil Masood and Naeem, Muhammad and Rashid, Imran},
  journal={IEEE Access}, 
  title={Energy Efficient Resource Allocation for {H-NOMA} Assisted {B5G} {HetNets}}, 
  year={2022},
  volume={10},
  number={},
  pages={91699-91711},}

@misc{cvx,
  author       = {CVX Research, Inc.},
  title        = {{CVX}: Matlab Software for Disciplined Convex Programming, version 2.0},
  howpublished = {\url{https://cvxr.com/cvx}},
  month        = aug,
  year         = 2012
}

@incollection{gb08,
  author    = {M. Grant and S. Boyd},
  title     = {Graph implementations for nonsmooth convex programs},
  booktitle = {Recent Advances in Learning and Control},
  series    = {Lecture Notes in Control and Information Sciences},
  editor    = {V. Blondel and S. Boyd and H. Kimura},
  publisher = {Springer-Verlag Limited},
  pages     = {95--110},
  year      = {2008},
}

@ARTICLE{DRLCellFreeDist,
  author={Fredj, Firas and Al-Eryani, Yasser and Maghsudi, Setareh and Akrout, Mohamed and Hossain, Ekram},
  journal={IEEE Trans. Cogn. Commun. Netw.}, 
  title={Distributed Beamforming Techniques for Cell-Free Wireless Networks Using Deep Reinforcement Learning}, 
  year={2022},
month={Jun.},
  volume={8},
  number={2},
  pages={1186-1201},
  doi={10.1109/TCCN.2022.3165810}}

@ARTICLE{GraphUABF,
  author={Lim, Byungju and Vu, Mai},
  journal={IEEE Trans. Wireless Commun.}, 
  title={Distributed Graph-Based Learning for User Association and Beamforming Design in Multi-{RIS} Multi-Cell Networks}, 
  year={2025},
month={Jul.},
  volume={24},
  number={7},
  pages={6118-6134},
  doi={10.1109/TWC.2025.3551763}}

@inproceedings{ADMMCellFree,
   title={{ADMM} for Downlink Beamforming in Cell-Free Massive {MIMO} Systems},
   DOI={10.1109/ieeeconf60004.2024.10943106},
   booktitle={2024 58th Asilomar Conference on Signals, Systems, and Computers},
   publisher={IEEE},
   author={Zafari, Mehdi and Pandey, Divyanshu and Doost-Mohammady, Rahman and Uribe, César A.},
   year={2024},
   month=oct, pages={623–628} }

@ARTICLE{ADMMforDist,
  author={Maros, Marie and Jaldén, Joakim},
  journal={IEEE Trans. Signal Inf. Process. Netw.}, 
  title={{ADMM} for Distributed Dynamic Beamforming}, 
  year={2018},
month={Jun.},
  volume={4},
  number={2},
  pages={220-235},
  doi={10.1109/TSIPN.2017.2681205}}

@misc{ADMMVariants,
      title={A Survey of {ADMM} Variants for Distributed Optimization: Problems, Algorithms and Features}, 
      author={Yu Yang and Xiaohong Guan and Qing-Shan Jia and Liang Yu and Bolun Xu and Costas J. Spanos},
      year={2022},
      eprint={2208.03700},
      archivePrefix={arXiv},
      primaryClass={cs.DC},
      url={https://arxiv.org/abs/2208.03700}, 
}

@ARTICLE{NTNApplications,
  author={He, Peng and Lei, Hailong and Wu, Dapeng and Wang, Ruyan and Cui, Yaping and Zhu, Yan and Ying, Zhaopeng},
  journal={IEEE Internet Things J.}, 
  title={Nonterrestrial Network Technologies: Applications and Future Prospects}, 
  year={2025},
  month={Mar.},
  volume={12},
  number={6},
  pages={6275-6299}}

@ARTICLE{InterferenceManagement,
  author={Siddiqui, Maraj Uddin Ahmed and Qamar, Faizan and Ahmed, Faisal and Nguyen, Quang Ngoc and Hassan, Rosilah},
  journal={IEEE Access}, 
  title={Interference Management in {5G} and Beyond Network: Requirements, Challenges and Future Directions}, 
  year={2021},
  volume={9},
  number={},
  pages={68932-68965},
  doi={10.1109/ACCESS.2021.3073543}}

@INPROCEEDINGS{BFforIM,
  author={Manai, Hosni and Ben Hadj Slama, Larbi and Bouallegue, Ridha},
  booktitle={2019 15th International Wireless Communications \& Mobile Computing Conference (IWCMC)}, 
  title={Interference Management by Adaptive Beamforming Algorithm in Massive {MIMO} Networks}, 
  year={2019},
  volume={},
  number={},
  pages={49-54}}

@ARTICLE{HuseyinArslan,
  author={Kirik, Muhammet and Alkana’neh, Aseel and Afeef, Liza and Arslan, HÜSeyin},
  journal={ IEEE Open J. Commun. Soc.}, 
  title={Efficient Interference Management Design for {NTN/TN} Co-Existence in {HAP}-Based {6G} Networks}, 
  year={2025},
  volume={6},
  number={},
  pages={5434-5449}}

@ARTICLE{HetNetBF,
  author={Xu, Yongjun and Xie, Hao and Hu, Rose Qingyang},
  journal={IEEE Commun. Lett.}, 
  title={Max-Min Beamforming Design for Heterogeneous Networks With Hardware Impairments}, 
  year={2021},
month={Apr,},
  volume={25},
  number={4},
  pages={1328-1332},
  doi={10.1109/LCOMM.2020.3044936}}

@ARTICLE{REBF,
  author={Han, Leixin and Wang, Jiaheng and Hou, Ruiding and He, Shiwen and Ng, Derrick Wing Kwan and Xia, Liang and Wang, Qixing},
  journal={IEEE Trans. Commun.}, 
  title={Resource Efficient Beamforming Design for Cell-Free Networks}, 
  year={2024},
  month={Dec.},
  volume={72},
  number={12},
  pages={7511-7525},
  doi={10.1109/TCOMM.2024.3418893}}

@ARTICLE{SpecSharing,
  author={Lee, Hao-Wei and Medles, Abdelkader and Chen, Chun-Chia and Wei, Hung-Yu},
  journal={IEEE Wireless Commun.}, 
  title={Feasibility and Opportunities of Terrestrial Network and Non-Terrestrial Network Spectrum Sharing}, 
  year={Dec. 2023},
  volume={30},
  number={6},
  pages={36-42}}

@article{SCA,
  author = {A. Beck and A. Ben-Tal and L. Tetruashvili},
  title = {A sequential parametric convex approximation method with applications to nonconvex truss topology design problems},
  journal = {Journal of Global Optimization},
  volume = {47},
  number = {1},
  pages = {29--51},
  year = {2010},
  month = {May}
}

@ARTICLE{PF,
  author={Diamantoulakis, Panagiotis D. and Karagiannidis, George K.},
  journal={IEEE Wireless Commun. Lett.}, 
  title={Maximizing Proportional Fairness in Wireless Powered Communications}, 
  year={2017},
  volume={6},
month={Apr.},
  number={2},
  pages={202-205},
  doi={10.1109/LWC.2017.2659729}}

@TechReport{itu_wrc23,

	institution    ={International Telecommunication Union (ITU)},
	title         = {{World Radiocommunication Conference 2023 (WRC-23) Final Acts}},
	year          = {2023},
	url		= {https://www.itu.int/wrc-23/},
	month         = {December},
}

@article{PFCharging,
  author    = {Frank Kelly},
  title     = {Charging and Rate Control for Elastic Traffic},
  journal   = {European Trans. Telecommun.},
  volume    = {8},
  number    = {1},
  pages     = {33--37},
  year      = {1997},
  doi       = {10.1002/ett.4460080106}
}

@ARTICLE{Ngo2017CellFree,
  author={Ngo, Hien Quoc and Ashikhmin, Alexei and Yang, Hong and Larsson, Erik G. and Marzetta, Thomas L.},
  journal={IEEE Trans. Wireless Commun.}, 
  title={Cell-Free Massive {MIMO} Versus Small Cells}, 
  year={2017},
  month={Mar.},
  volume={16},
  number={3},
  pages={1834-1850},
  doi={10.1109/TWC.2017.2655515}}

@ARTICLE{APS_BF2026,
author={Khoshkbari, Hesam and Kaddoum, Georges and Abbasi, Omid and Selim, Bassant and Yanikomeroglu, Halim}, journal={IEEE Trans. Commun.}, 
title={Beamforming for Massive {MIMO} Aerial Communications: A Robust and Scalable {DRL} Approach}, 
year={2026}, 
volume={74},
number={},
pages={261-275}}

@ARTICLE{ZiyeJia,
  author={Jia, Ziye and Sheng, Min and Li, Jiandong and Zhou, Di and Han, Zhu},
  journal={IEEE J. Sel. Areas Commun.}, 
  title={Joint {HAP} Access and {LEO} Satellite Backhaul in {6G}: Matching Game-Based Approaches}, 
  year={2021},
  month={Apr.},
  volume={39},
  number={4},
  pages={1147-1159}}

@ARTICLE{ZiyeMEC,
  author={Jia, Ziye and Cui, Can and Dong, Chao and Wu, Qihui and Ling, Zhuang and Niyato, Dusit and Han, Zhu},
  journal={IEEE Trans. Mob. Comput.}, 
  title={Distributionally Robust Optimization for Aerial Multi-Access Edge Computing via Cooperation of {UAVs} and {HAPs}}, 
  year={2025},
  month={Oct.},
  volume={24},
  number={10},
  pages={10853-10867}}

@ARTICLE{nuImpact,
  author={El Haber, Elie and Nguyen, Tri Minh and Assi, Chadi},
  journal={IEEE Trans. Commun.}, 
  title={Joint Optimization of Computational Cost and Devices Energy for Task Offloading in Multi-Tier Edge-Clouds}, 
  year={2019},
  volume={67},
  month={May},
  number={5},
  pages={3407-3421}}
 \appendices
 \color{black}
\section{Tightness of the Reformulated Constraints in Section~\ref{Sec:Reformulation}}\label{Appendix_A}

In this appendix, we prove that all auxiliary-variable inequalities introduced in the distributed reformulation hold with equality at the optimal solution. Furthermore, we show that the Jensen-based approximation is tight in the cell-free network. This can be proved by contradiction.
The objective function $f(\mathbf{t})=\sum_{u}\log(t_u)$ is monotonically increasing in each $t_u>0$. Therefore, the optimizer always selects $e^{t_u^\star} = 1+\alpha_u^\star$, which makes constraint \eqref{exp_const} active at optimality. Otherwise, there exists $\hat{t}_u$ such that $e^{t_u^\star} < e^{\hat{t}_u} \leq 1+\alpha_u^\star$, yielding a strictly larger objective value, which contradicts optimality.
Similarly, larger values of $\alpha_u$ enlarge the feasible region of $t_u$ and improve the objective function value. Hence, constraint \eqref{eq:alphabetaileqPRiV2} must also be active at optimality. Moreover, since increasing $\beta_u$ makes constraint \eqref{eq:alphabetaileqPRiV2} more restrictive, the optimizer always selects the minimum feasible $\beta_u$, implying that constraint \eqref{eq:betaigeqIiV2} is active at optimality. Likewise, reducing any $I_{b,u}$ decreases $\beta_u$ and enlarges the feasible region of $\alpha_u$, which means that constraint \eqref{eq:PF_Const1_new2} is also active at optimality.

Finally, the Jensen-based reformulation in \eqref{Jensen's Inequality} introduces the approximation gap
\begin{equation}
\small
\delta_u \triangleq
\sum_{b\in \mathcal{B}}\sum_{k \in \mathcal{U} \setminus \{u\}}
{\left|\left(\mathbf{h}^{b}_{u}\right)^H \mathbf{w}^{b}_{k}\right|^2}
-
\sum_{k \in \mathcal{U} \setminus \{u\}}
\left|
\sum_{b\in \mathcal{B}}
{\left(\mathbf{h}^{b}_{u}\right)^H \mathbf{w}^{b}_{k}}
\right|^2.
\end{equation}

Accordingly, expanding the second term yields
\begin{equation}
\small
\delta_u
=
-\sum_{k \in \mathcal{U} \setminus \{u\}}\sum_{\substack{b,b' \in \mathcal{B}\\ b \neq b'}}
\left(\mathbf{h}^{b}_{u}\right)^H \mathbf{w}^{b}_{k}
\left(
\left(\mathbf{h}^{b'}_{u}\right)^H \mathbf{w}^{b'}_{k}
\right)^*.
\end{equation}

Since channel vectors $\mathbf{h}_u^b,\forall b$ are independent, the expectation with respect to channel on $\delta_u$ is negligible, i.e.,:
\begin{equation}
    \mathbb{E}\!\left[
\left(\mathbf{h}^{b}_{u}\right)^H \mathbf{w}^{b}_{k}\left(\left(\mathbf{h}^{b'}_{u}\right)^H \mathbf{w}^{b'}_{k}
\right)^*\right] = 0,~\forall b,b' \in \mathcal{B},~b\neq b'.
\end{equation}

As a result, the Jensen-based approximation, on average, is tight.

\color{black}
\vfill

\end{document}